\documentclass[myepj-spec,pdftex]{mySvjour}
\usepackage{graphicx}
\usepackage[pdfpagemode=UseNone]{hyperref}
\usepackage{color}
\usepackage{xspace}
\usepackage[square,sort&compress,numbers]{natbib}   
\usepackage{verbatim}
\usepackage{amsmath,amssymb,amsfonts} 
\usepackage{tabularx}
\usepackage{multicol}
\usepackage{footnote}
\usepackage{listings}
\usepackage[gen]{eurosym}
\usepackage[switch, modulo]{lineno}

\usepackage{lmodern,bm}                
\usepackage[T1]{sansmath} 
\SetMathAlphabet{\mathsfbf}{sans}{\sansmathencoding}{\sfdefault}{bx}{sl}
\usepackage{etoolbox}
\AtBeginEnvironment{sansmath}{}{}{}

\usepackage{lipsum}

\newcommand{\pT}{\ensuremath{p_{\mathrm{T}}}}

\newcommand{\ET}{\ensuremath{E_{\mathrm{T}}}}
\newcommand{\MET}{\mbox{\ensuremath{\not \!\! \ET}}}

\definecolor{darkblue1}{rgb}{0,0,.2}
\definecolor{darkblue}{rgb}{0,0,.2}
\definecolor{darkred}{rgb}{0.5,0,0}
\pagecolor{white} 
\color{black}     
\hypersetup{breaklinks=true, 
	colorlinks=true, 
	linkcolor=darkblue1, 
	menucolor=darkblue1, 
	urlcolor=darkblue1,
	citecolor=darkblue1,
	pdftitle={},
	pdfauthor={},
	pdfsubject={},
	pdfkeywords={},
	pdfproducer={}
}
\usepackage{siunitx}

\bibstyle{plain}
\newcommand{\bi}{\begin{itemize}}
\newcommand{\ei}{\end{itemize}}

\newcommand{\ben}{\begin{enumerate}}
\newcommand{\een}{\end{enumerate}} 

\newcommand{\bt}[1]{\begin{table}[tb]\begin{tabular}{#1} \hline\hline  \\[-1.0em]}
\newcommand{\et}[2]{\hline\hline \end{tabular} \caption{#1} \label{#2} \end{table}}

\newcommand{\be}{\begin{equation}}
\newcommand{\ee}{\end{equation}}

\newcommand{\bea}{\begin{eqnarray}}
\newcommand{\eea}{\end{eqnarray}}

\newcommand{\bc}{}

\newcommand{\mev}{\ensuremath{\mathrm{\,Me\kern -0.1em V}}\xspace}
\newcommand{\gev}{\ensuremath{\mathrm{\,Ge\kern -0.1em V}}\xspace}

\begin{document}
			
			\begin{flushright}
				\normalsize
			\end{flushright}
			
			\vspace{-2cm}
			
			\title{\Large\boldmath Prospects of Searches for Anomalous Hadronic Higgs Boson Decays at the LHeC}
			%

\author{Subhasish Behera \and Manuel Hagelüken \and Matthias Schott}
\institute{Institute of Physics, Johannes Gutenberg University, Mainz, Germany}

			
			\abstract{%
The future Large Hadron Electron Collider (LHeC) would enable collisions of an intense electron beam with either protons or heavy ions at the High Luminosity-Large Hadron Collider (HL-LHC). With a center of mass energy greater than a TeV and a high luminosity, the LHeC would be a new-generation collider for deep-inelastic scattering (DIS) and a significant facility for precise Higgs physics, complementing $pp$ and electron-positron colliders. Anomalous hadronic decay signatures of the Higgs Boson, such as those into three or more partons, are challenging to detect at the LHC due to high background rates, but they may be observable at the LHeC. This paper presents the expected sensitivity of the LHeC for the decay channels $H\rightarrow 3jets$ and $H\rightarrow 4jets$, assuming an integrated luminosity of 1 ab$^{-1}$. Our analysis indicates that upper limits on the branching ratio of the Higgs Boson of 0.35 and 0.17 at 95\% confidence limit for these processes are possible.}
	\maketitle


\section{Introduction}	

A future Large Hadron Electron Collider (LHeC) \cite{LHeC:2020van, LHeCStudyGroup:2012zhm} at CERN would collide $7$ TeV LHC proton beams and $60$ GeV electron beams at a luminosity of $10^{34}~\text{cm}^{-2}\text{ s}^{-1}$. This would take place in parallel to the proton-proton collisions of the LHC. The design for this electron accelerator is based on a linac-ring $ep$ collider configuration with two superconducting linacs, each below 1km in length, operating in continuous wave (CW) energy recovery mode \cite{Bartnik:2020pos}. The electron beam can have an effective left/right polarization up to 80\%. The main focus of the LHeC physics program is deep inelastic scattering (DIS) physics, probing a completely new area of the low-$x$ phase space, allowing the precise determination of proton and nuclear parton distribution functions (PDFs) \cite{Lai:2010vv}. PDFs are an essential pre-requirement for any future high-energy hadron collider(s), including LHC. In addition to the DIS program, searches for physics beyond the Standard Model, such as leptoquarks, contact interactions, as well as RPV and SUSY promise significantly higher sensitivities than is currently possible at existing colliders. The LHeC also has significant potential for measurements as well as searches in the Higgs sector. Here, Higgs boson production via vector-boson-fusion and its decay into $b$- and $c$- quarks could be much cleaner than at the LHC \cite{Han:2009pe}, allowing us to probe the relevant Higgs couplings to a higher precision~\cite{Behera:2022gnr, LHeCStudyGroup:2012zhm}. 

Most of the direct standard model decay channels of the Higgs Boson, in particular into $WW$, $ZZ$, $bb$, $\tau\tau$ and $\mu\mu$ have been or can be probed at the LHC to a high precision \cite{Yang:2020omr}. One possibility to probe all possible Higgs boson decays is the direct measurement of its full decay width. Since Higgs width measurements at the LHC via off-shell production \cite{ATLAS:2022rks, CMS:2022ley} incur a significant model dependence, a full model-independent measurement can only be performed at an electron-positron collider with a center of mass energy above the Higgs production threshold \cite{Behnke:2013xla}. Hence, a large number of direct searches for anomalous Higgs Boson decays have been conducted at the LHC (e.g. \cite{CMS:2021nnc, ATLAS:2014kca}), in particular, the Higgs to invisible particle decays \cite{ATLAS:2023tkt}, yielding a limit about 0.1. Notoriously challenging are searches for anomalous hadronic Higgs Boson decays at the LHC, due to the enormous background rates of multi-jet processes. Here, the Higgs boson production and decay at the LHeC might offer promising options, since the Higgs boson production processes do not involve any colored initial states but stem solely from vector boson fusion processes, and at the same time, the multi-jet final state background processes are significantly reduced. In this paper, we estimate the sensitivity of searches for anomalous Higgs boson decays in three and four partons at the LHeC for the first time, assuming an integrated luminosity of 1 ab$^{-1}$. It should be noted that this study should serve only as a starting point for other direct searches for anomalous Higgs boson decays in more generic hadronic final states.

\section{Monte Carlo Simulations\label{sec:MC}}

The proposed baseline energy of the electron beam is $60$ GeV, which in combination with the $7$ TeV proton beam results in a center of mass energy of $1.3$ TeV. This is about four times that of its predecessor, HERA~\cite{Wiik:1985sb} at DESY, which had a center of mass energy of $318$ GeV. In addition, the expected luminosity is about three orders of magnitude higher.\par
\begin{figure*}[thb]
\centering
    \includegraphics[width=0.32\textwidth]{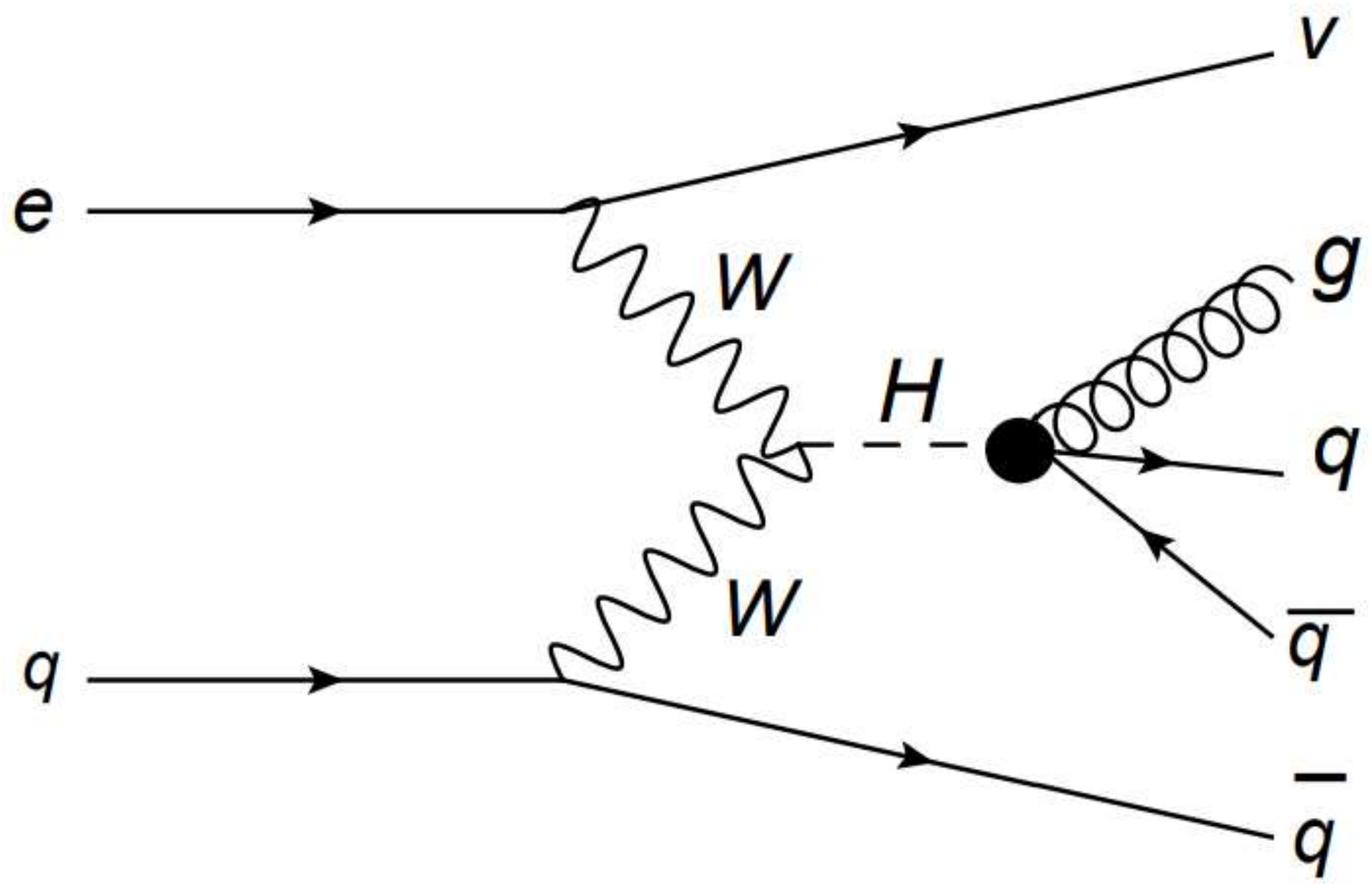}
\includegraphics[width=0.32\textwidth]{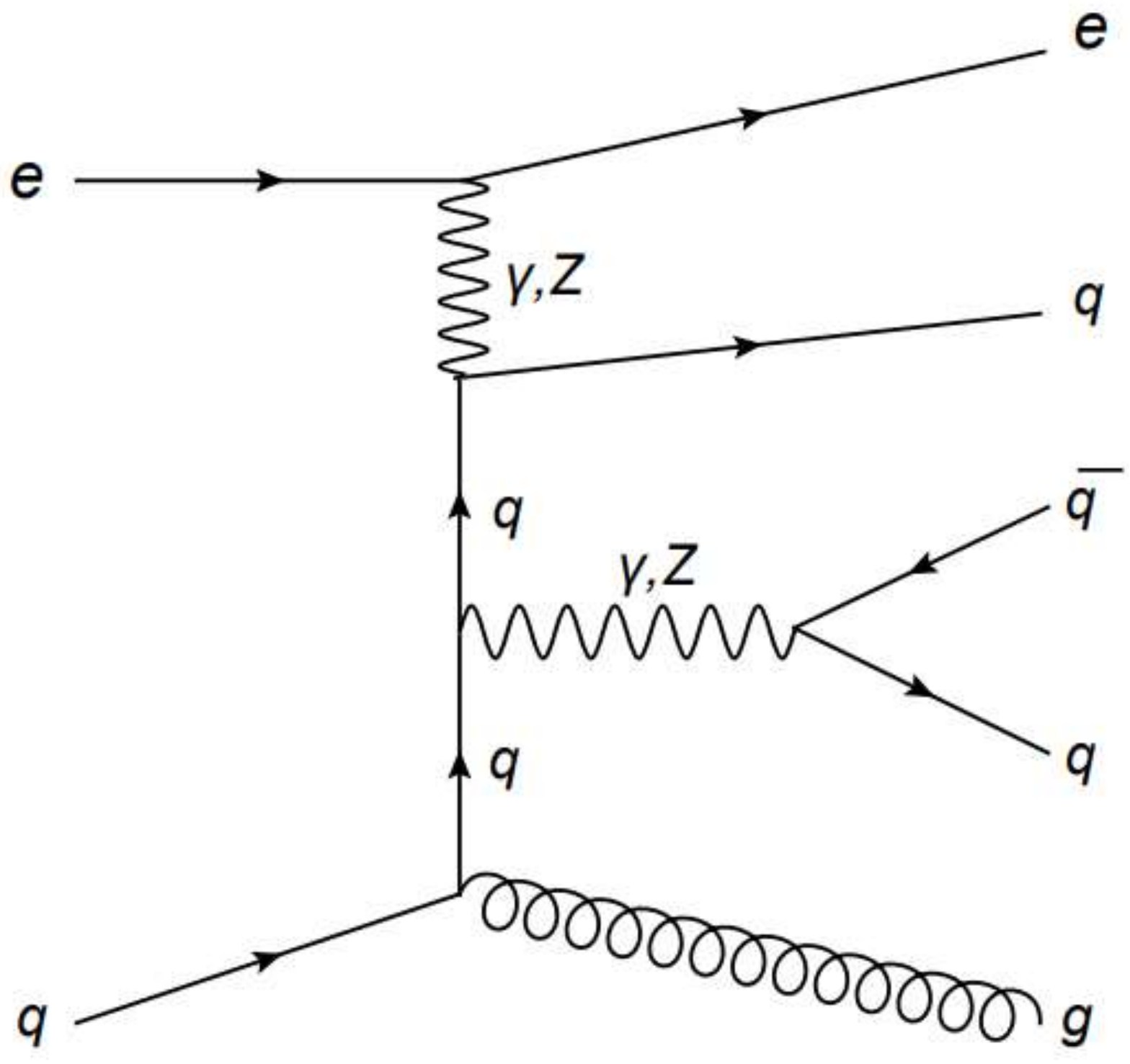}
\includegraphics[width=0.32\textwidth]{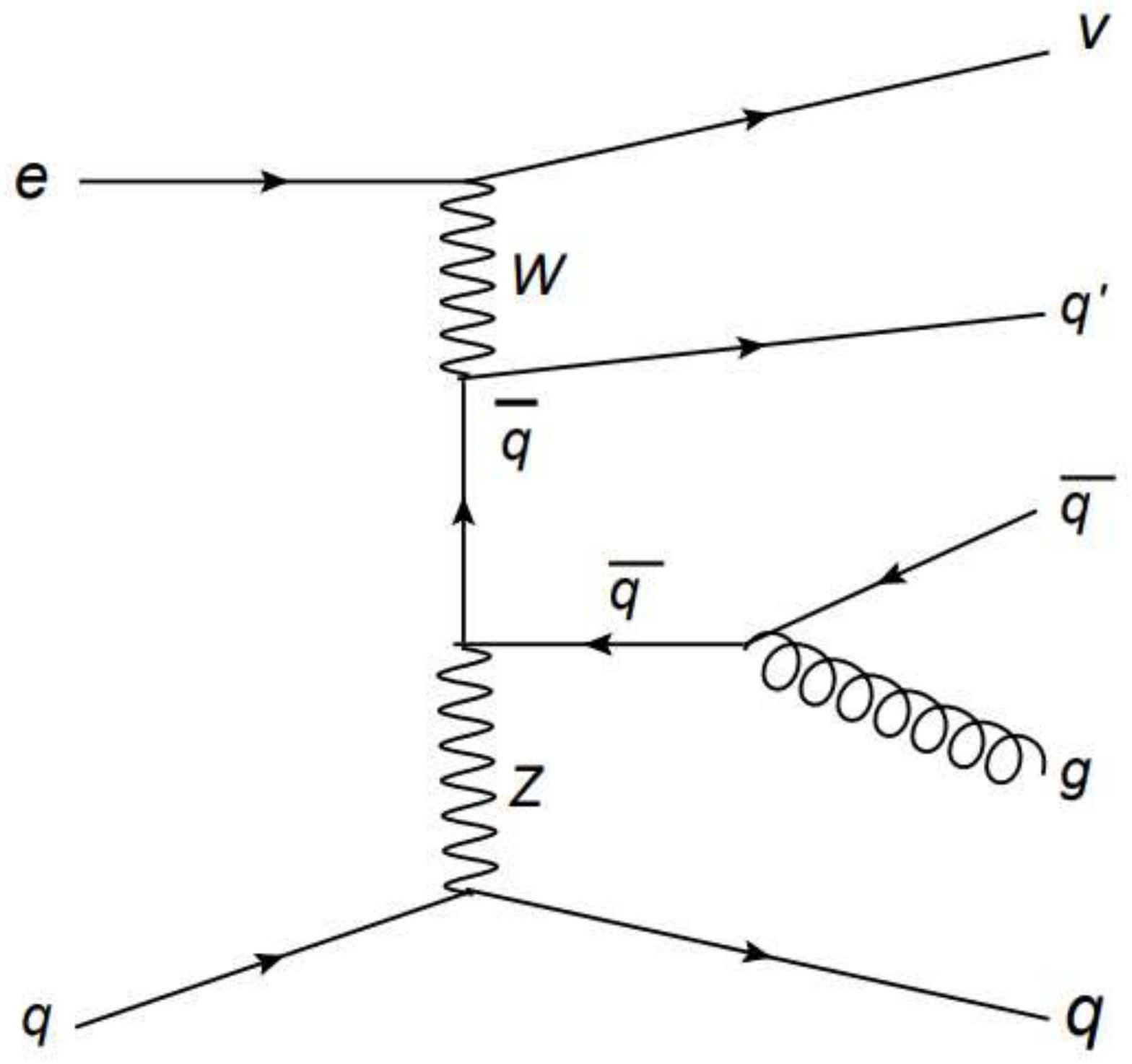}
    \includegraphics[width=0.32\textwidth]{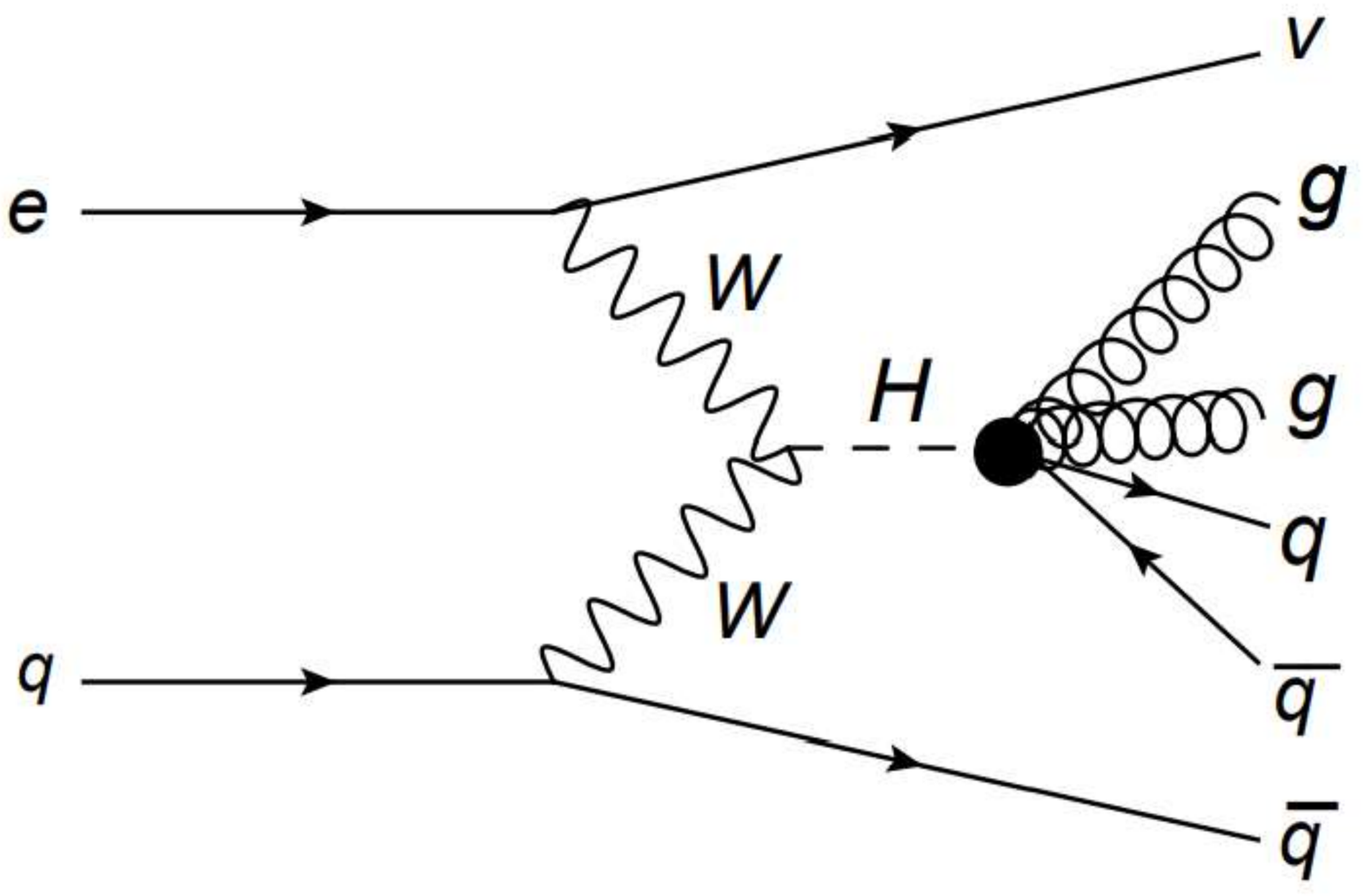}
\includegraphics[width=0.32\textwidth]{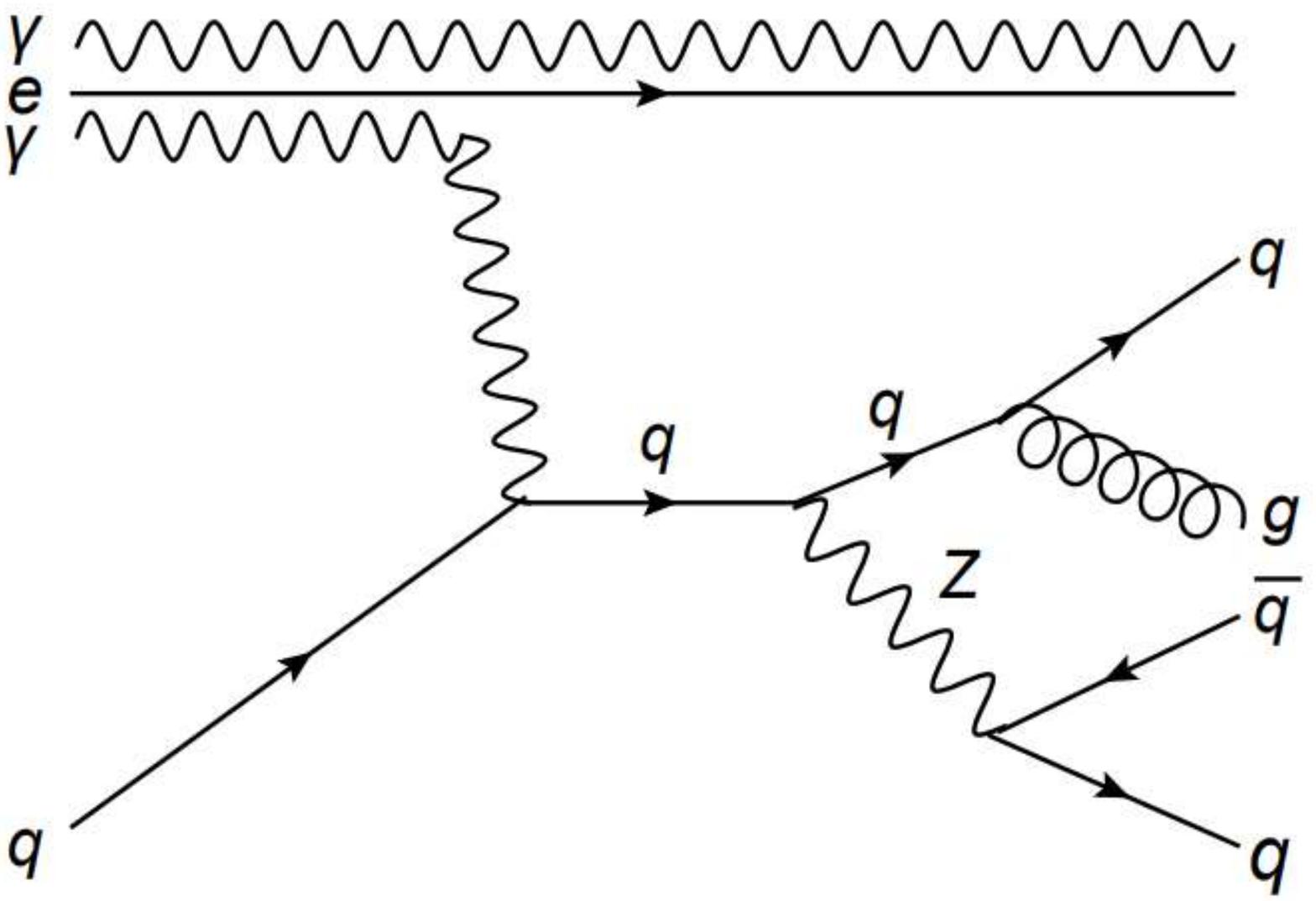}
\includegraphics[width=0.32\textwidth]{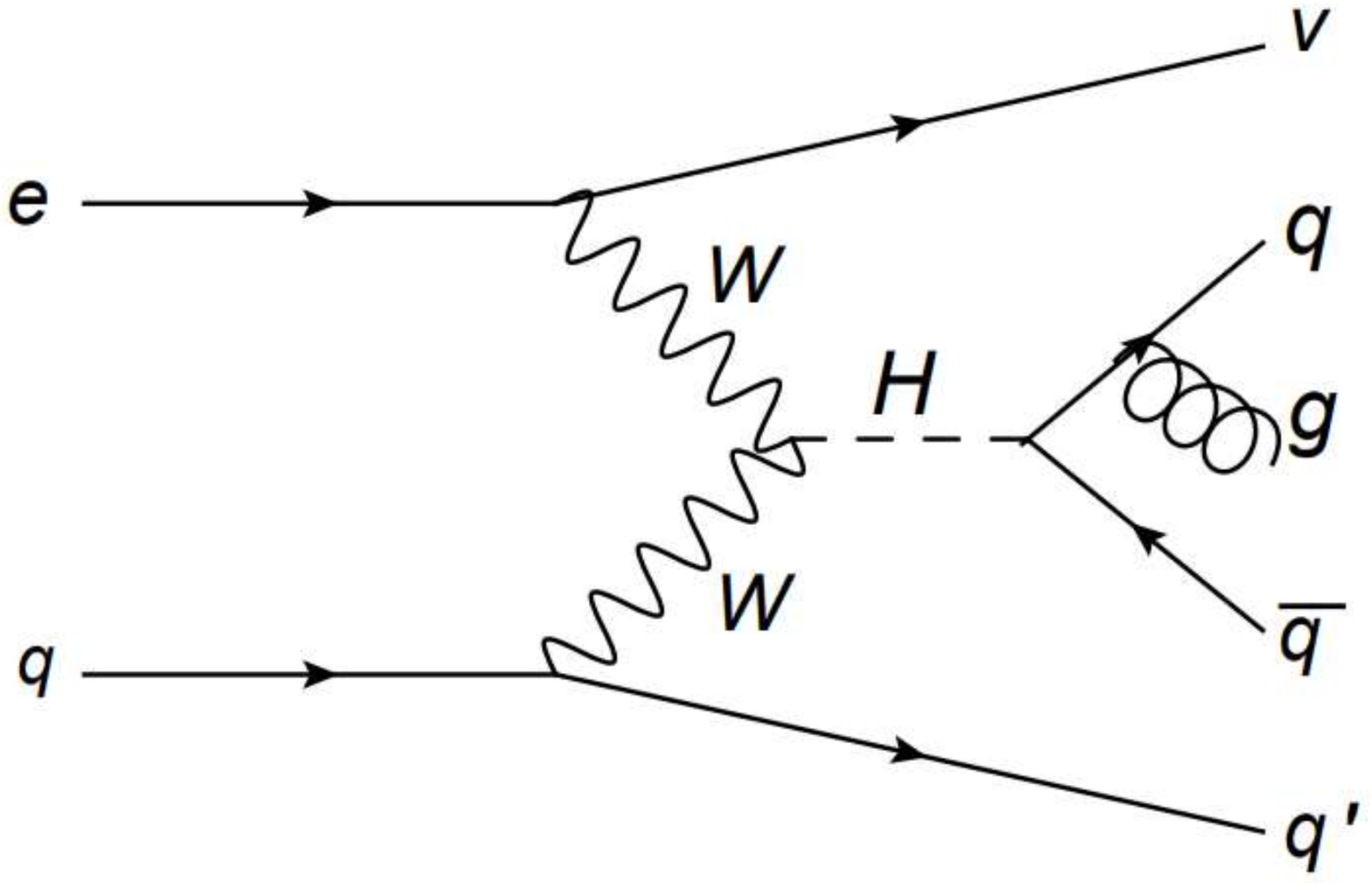}
    \caption{Feynman Diagram of selected signal (left) and background processes (middle, right).}
\label{fig:feynman}
\end{figure*}

Higgs bosons in $ep$ collisions will be produced through vector boson fusion via either a charged current (CC) or a neutral current (NC) interaction. Since the production cross section for charged current interactions is dominant for left polarised initial state electron, we focus only on the associated final state of $\nu_e X j$ in this work, where $j$ represents the light quark jet, termed as the leading jet with the highest $p_T$ jet in the analysis. The signal processes are anomalous Higgs boson decays into three and four partons (quarks and gluons) depicted on the left of Figure \ref{fig:feynman}. The considered background processes are also shown schematically in Figure \ref{fig:feynman} and can be distinguished between charged current (CC) and neutral current (NC) induced processes, photo-production of multi-jet final states (Photo), as well as Higgs boson decays into quarks with an additional parton (H+Jets). 

The \textsc{MadGraph5} generator \cite{Alwall:2014hca} has been developed to model the hard scattering of proton-electron collisions for all relevant signal and background processes, using the CTEQ6L1 PDF set \cite{Pumplin:2002vw}. 

 The signal is generated using the Higgs effective Lagrangian containing the operators up to dimension six allowing three- and four-point interactions involving Higgs and gauge bosons \cite{Alloul:2013naa}. The Universal Feynman Output (UFO) file is developed using the publicly available tools {\sc Feynrules 2.0} \cite{Christensen:2008py, Alloul:2013bka}. The UFO is also available publicly from the {\sc Feynrules 2.0} database \cite{UFOHEL}. We generate the signal samples by switching on the $H \rightarrow 3j$ and $H \rightarrow 4j$ couplings only. The relevant part of the Lagrangian for the signal process is,
\begin{equation}
    {\cal L} \supset - \frac{2\sqrt{2}g_s}{m^2_W} y_q \bar{C}_{\rm qG}[\bar{q} T_a \gamma^{\mu \nu} P_R q] G^a_{\mu \nu}h + h.c.,
\end{equation}
where the q in the Lagrangian represents either an up- or down-type quark but never the top quark. The $\bar{C}_{\rm qG}$ is the anomalous coupling constant. Other notations in the Lagrangian carry their usual meaning. The signal processes are generated at a fixed renormalization ($\mu_R$) and factorization ($\mu_F$) scale, equal to the $m_H=125~GeV$ with a coupling value set to $0.1$, whereas the background processes are generated without setting a fixed scale. For the generation of signals and backgrounds samples, we keep the initial electron beam
polarization to be 80\% left polarised.
The showering and hadronisation of the hard scattering events were carried out using \textsc{Pythia8.303} \cite{Sjostrand:2014zea}. To control the cross-section of the background processes during the event generation, several requirements on the transverse momentum, $\pT$, pseudo-rapidity, $\eta$, of the final charged leptons and quarks and on the invariant mass of two final state quarks as well as the total hadronic transverse momentum, $H_T$ in the final state have been applied (See Table \ref{tab:samples}). 
\par
At least 100k events for all signal and background samples have been produced, in order to get sufficient statistics after the final selection. A summary of all generated samples, including the applied generator-level cuts and the corresponding cross-section predictions, are summarised in \autoref{tab:samples}.

Once all samples for the signal and background processes are available on the generator level, the detector response for a future LHeC detector has been simulated using the  \textsc{Delphes} framework~\cite{deFavereau:2013fsa}, based on a dedicated configuration file \cite{LHeCDelphesCard}. The \textsc{Delphes} setup card specific for LHeC fast simulation was developed by the LHeC working group. While the effect of triggers at the LHeC was not studied here explicitly, we argue that dedicated triggers will have a sufficiently high efficiency so that the impact on the subsequent analysis is minimal. However, we do not assume any efficiency on a dedicated forward electron tagger, which would  allow suppression of all NC induced processes.

\begin{table}[ht]
\scriptsize
\begin{tabularx}{\linewidth}{l | l  | l | c | c | c }
\hline
Description 			&	Process								& Generator-Level Cuts	 			& Generator			& Cross Section	& Number Of	Events\\
    & & & & [pb] & $\times10$ \\
\hline
Signal ($H>3j$) &$p e^- \to \nu_e h(h\to 3q) q$& $|\eta| <10, ~\pT^q>10~GeV$	& \begin{tabular}{@{}c@{}}MadGraph5+ \\ Pythia8\end{tabular}&-&200 \\	
\hline
Signal ($H>4j$) &$p e^- \to \nu_e h(\to 4q) q$& $|\eta| <10, ~\pT^q>10~GeV$	& \begin{tabular}{@{}c@{}}MadGraph5+ \\ Pythia8\end{tabular}&-&200\\
\hline
CC & \begin{tabular}{@{}l@{}l@{}}$p e^- \to \nu_e 3q $ \\ $p e^- \to \nu_e 4q $ \\ $p e^- \to \nu_e 5q$ \end{tabular} & \begin{tabular}{@{}c@{}}$\pT^q>10~GeV,$ \\ $|\eta|<10,~50<H_T<400 ~GeV$\end{tabular}	& \begin{tabular}{@{}c@{}}MadGraph5+ \\ Pythia8\end{tabular} &\begin{tabular}{@{}c@{}c@{}} 51.1\\18.0 \\5.4\end{tabular} &\begin{tabular}{@{}c@{}c@{}} 200\\200\\200\end{tabular}	\\
\hline
H+Jets & \begin{tabular}{@{}l@{}l@{}}$p e^- \to \nu_e H(\to qq)q$ \\ $p e^- \to \nu_e H(\to qq)2q$ \\ $p e^- \to \nu_e H(\to qq)3q$ \end{tabular} & \begin{tabular}{@{}c@{}}$\pT^q>10~GeV,$ \\ $|\eta|<10,~50<H_T<400 ~GeV$\end{tabular}	& \begin{tabular}{@{}c@{}}MadGraph5+ \\ Pythia8\end{tabular} &\begin{tabular}{@{}c@{}c@{}} 0.133\\0.033 \\0.006\end{tabular} &\begin{tabular}{@{}c@{}c@{}} 200\\200\\200\end{tabular}	\\
\hline
NC&\begin{tabular}{@{}l@{}}$p e^- \to e^- 3q$ \\ $p e^- \to e^- 4q$\end{tabular}&\begin{tabular}{@{}c@{}}$\pT^q>10~GeV,$ \\ $|\eta|<10,~50<H_T<400 ~GeV$\end{tabular}	& \begin{tabular}{@{}c@{}}MadGraph5+ \\ Pythia8\end{tabular} &\begin{tabular}{@{}c@{}}1860 \\ 842\end{tabular}&\begin{tabular}{@{}c@{}}200 \\ 138\end{tabular}\\
\hline
Photo & \begin{tabular}{@{}l@{}l@{}}$p \gamma \to 3q$ \\ $p \gamma \to 4q$ \\ $p \gamma \to 5q$ \end{tabular} & \begin{tabular}{@{}c@{}}$\pT^q>10~GeV,$ \\  $|\eta|<10,~50<H_T<400 ~GeV$\end{tabular}	& \begin{tabular}{@{}c@{}}MadGraph5+ \\ Pythia8\end{tabular} &\begin{tabular}{@{}c@{}c@{}} 1658 \\ 580 \\ 110\end{tabular} &\begin{tabular}{@{}c@{}c@{}} 200\\200\\200\end{tabular}	\\
\hline
\end{tabularx}
\caption{The cross-section of the signal and all possible background samples for corresponding generator-level cuts are shown in the table. Here, $q$ represents either a quark or anti-quark of any SM flavor except top quark.\label{tab:samples}}
\end{table}

\section{Signal Selection and Background Estimation\label{sec:Signal}}

The selection of DIS-induced processes is improved through the application of several kinematic selection requirements to all events. The missing transverse energy, $\MET$, which is defined as the negative vector sum of all reconstructed cluster energies in the transverse plane, must be greater than 5 GeV. In order to suppress NC-induced background processes, no electron and muon candidates with a minimal transverse momentum of 5 GeV should be reconstructed. Furthermore, events must have at least one reconstructed jet in the forward region with $p_T>5$ GeV and $2.0<\eta<6.0$. The highest energetic jet among these jets is designated as the highest energetic forward jet and saved for further analysis. In addition, at least three or four jets with a minimal transverse momentum of $p_T>5$ GeV and $|\eta|<3$ must be found and considered for further analysis. These jets, excluding the highest energetic forward jet, are used to define the invariant signal jet mass, as they may originate from anomalous Higgs boson decays. Their invariant masses are denoted as $m_{j \geq 3}$ and $m_{j \geq 4}$, respectively.

\begin{figure}[t]
\centering
    \includegraphics[width=0.32\textwidth]{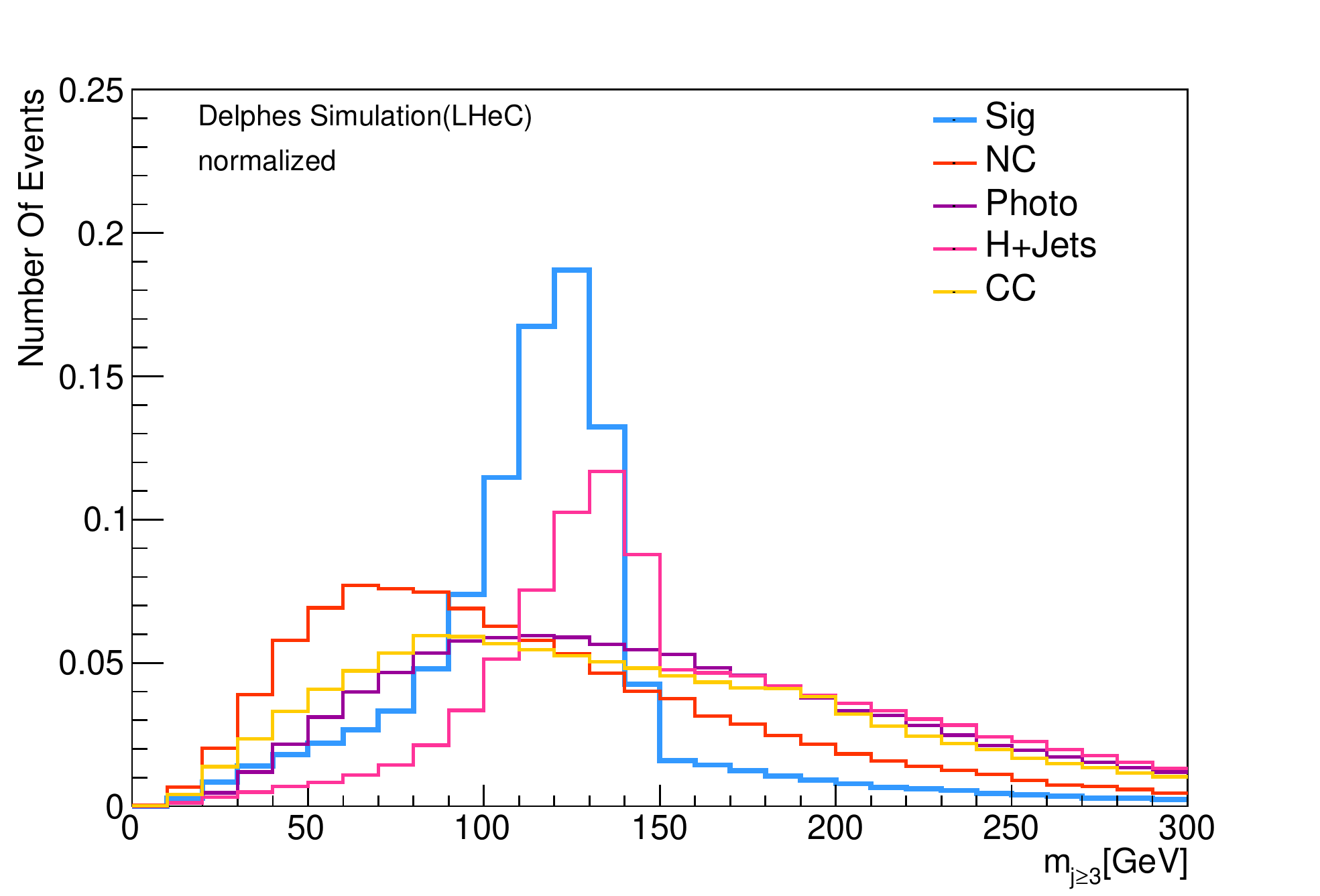}
    \includegraphics[width=0.32\textwidth]{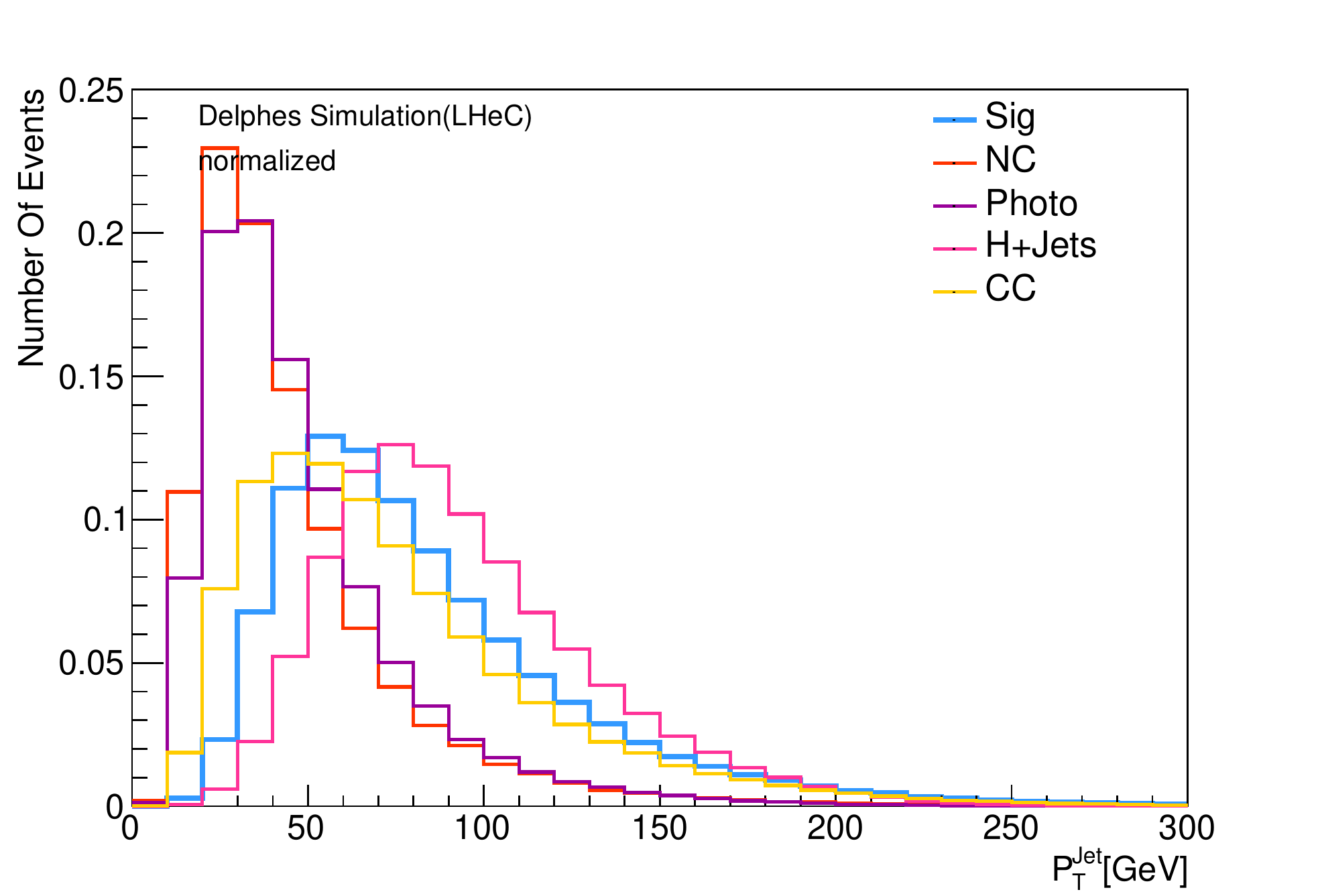}
    \includegraphics[width=0.32\textwidth]{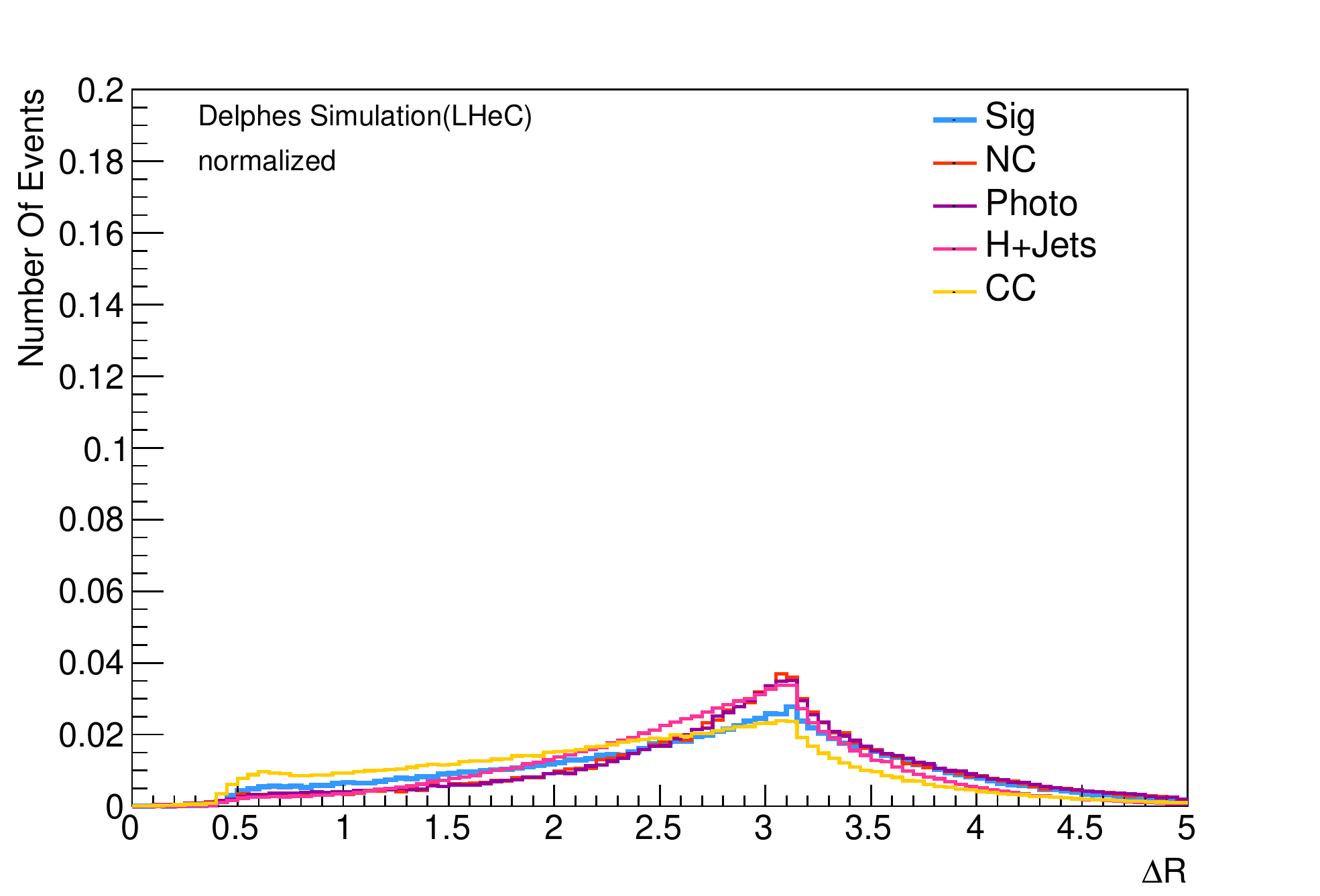}
    \includegraphics[width=0.32\textwidth]{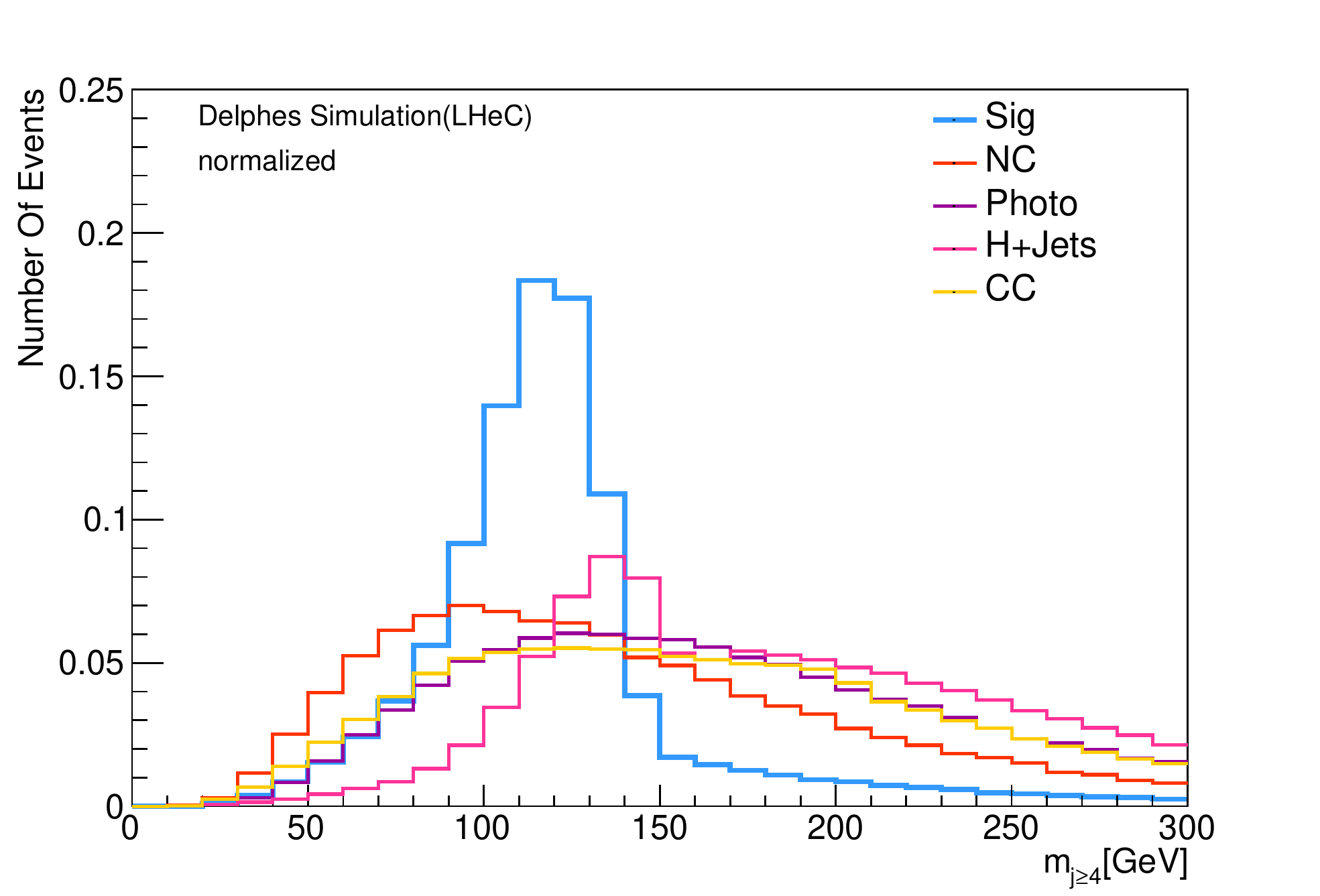}
    \includegraphics[width=0.32\textwidth]{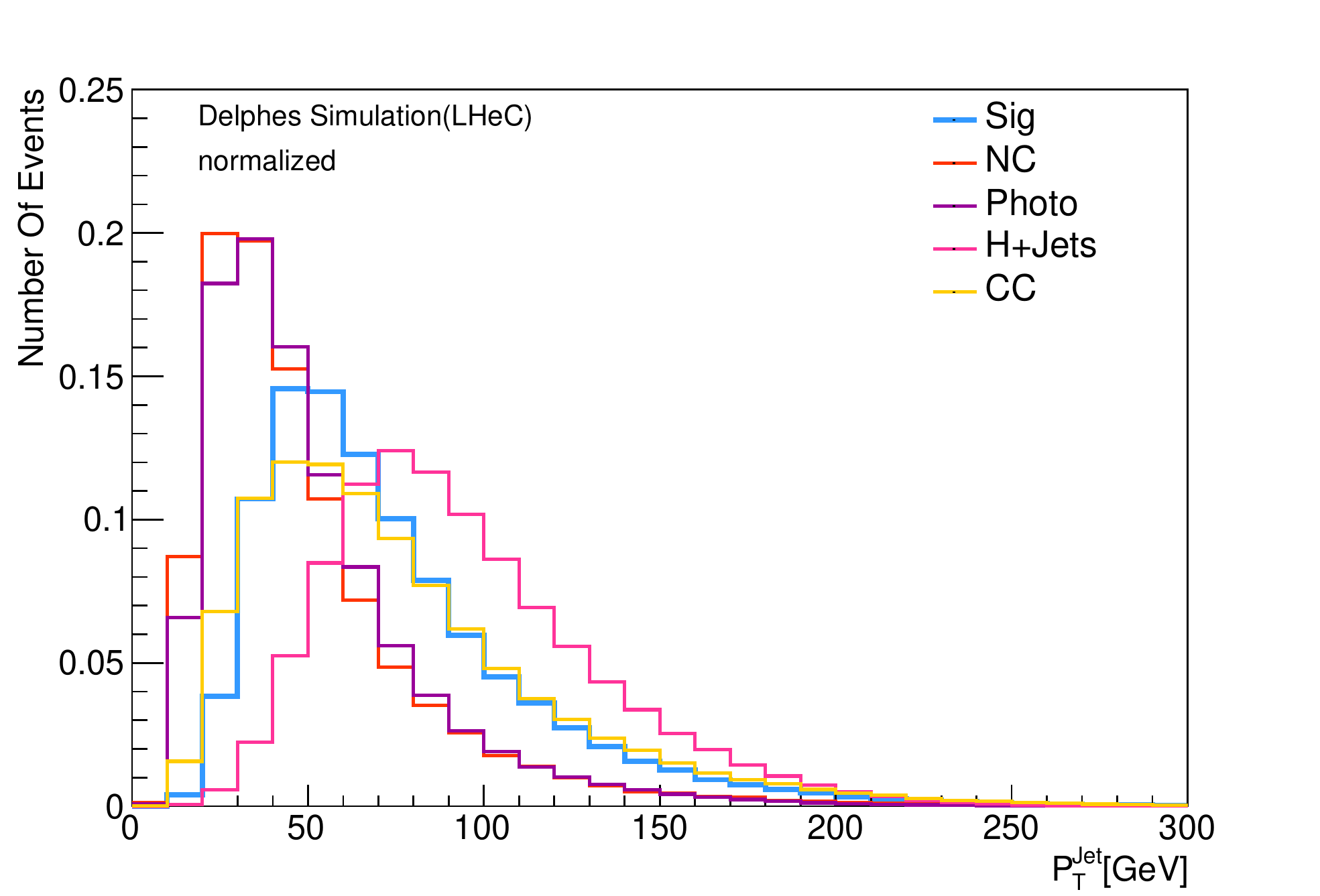}
    \includegraphics[width=0.32\textwidth]{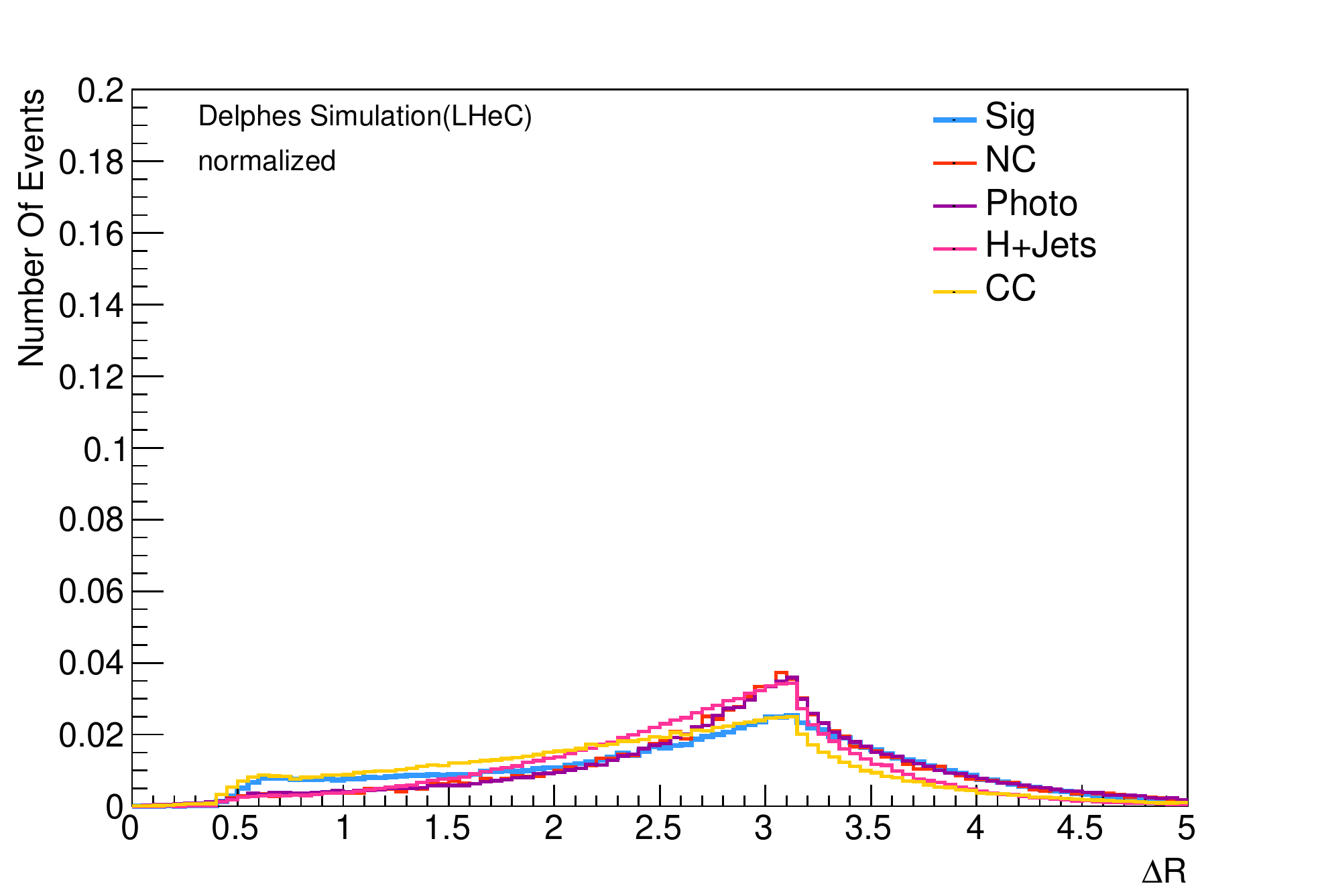}
    \caption{Normalized distributions of the invariant signal jet mass (left), the leading jet $p_T$ (middle) and the opening angle between two jets with leading $p_T$ (right) for $H\rightarrow 3j$ (upper row) and $H\rightarrow 4j$ (lower row) searches. The invariant signal jet mass is defined as the invariant mass of all jets passing the pre selection except the highest energetic forward jet as described in section 3.}
\label{fig:baseline34J}
\end{figure}

The normalized distributions of selected observables, such as the invariant signal jet masses, the leading jet $p_T$, and the opening angle between the two leading jets for signal and background processes in searches for $H\rightarrow3j$ and $H\rightarrow4j$, are depicted in Figure \ref{fig:baseline34J}. As anticipated, the signal peaks around the Higgs boson mass, with tails towards lower masses due to parton shower processes of Higgs decay partons, as well as a smaller tail towards higher masses due to false jet selections. Significant differences in the other kinematic observables are also evident for neural network based classifiers.

Neural network based classifiers are implemented to separate signal and background processes. In total, four networks for the separation of the NC, CC, photo-induced (Photo) and $H\rightarrow q\bar q$ (H+Jets) processes are used. The inputs to each network consist of kinematic variables of the twelve leading jets, including the highest energetic forward jet, namely transverse momentum ($p_T$), pseudo-rapidity ($\eta$), azimuthal angle ($\phi$), and mass ($m$), as well as whether or not the jet is b-tagged (for jets within $|\eta|<2.5$). The total number of selected jets, the scalar sum of all selected jet $p_T$'s, the missing transverse energy, and its $\phi$ direction are also considered as inputs, resulting in an input layer with 65 neurons. The network architecture consists of two hidden layers with 128 neurons each, and one output layer with a single neuron and a sigmoid activation function.

The neural network based classifiers have been implemented in the \textsc{Keras} framework \cite{chollet2015keras} and trained using \textsc{Tensorflow} \cite{abadi2016tensorflow}. All signal and background samples have been split into training (60\%), validation (20\%) and test-samples (20\%). The training is based on the \textsc{ABCDisCo} method \cite{Kasieczka:2020pil}, where the network does not only consider the separation between signal and background processes in the loss function $L$ but also the correlation to a given observable.  In our study, we chose to train the networks to decorrelate to the $m_{j \geq 3}$ and $m_{j \geq 4}$ distributions, respectively. The loss function can be generically expressed as $L = \sqrt{(CRE)^2 + (\alpha DisCo(m_{Xj},o))^2}$, where $CRE$ is the typical cross-entropy loss definition on the training sample, $DisCo$ is the distance correlation of the network output $o$ and the $m_{j \geq 3}$/$m_{j \geq 4}$ values of the background sample and $\alpha$ denotes the relative importance between $CRE$ and the correlation term.  While the required decorrelation negatively impacts the classification performance, it has two advantages: First of all, the sensitivity on the signal process can be extracted using a template fit on the $m_{j \geq 3}$ and $m_{j \geq 4}$ distributions, where a distinct signal peak is expected around the Higgs boson mass. The peak structure allows for a data-driven control of the background processes in the off-peak regions. Secondly, the background shapes stays invariant after the application of a cut on the network output, hence also small remaining statistics of background processes after the full selection can be used to estimate the background shape in detail.

\begin{figure}[thb]
\centering
    \includegraphics[width=0.49\textwidth]{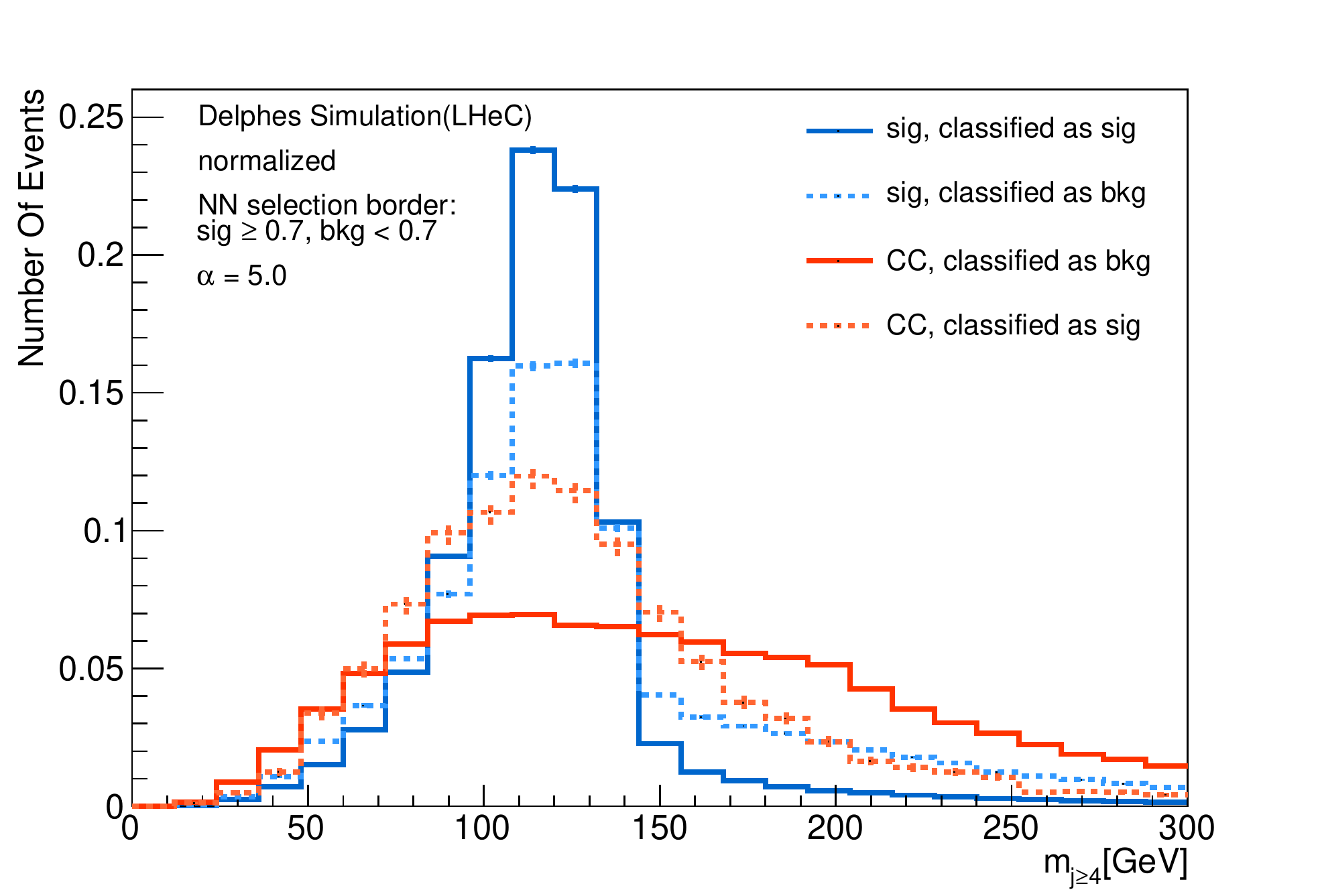}
    \includegraphics[width=0.49\textwidth]{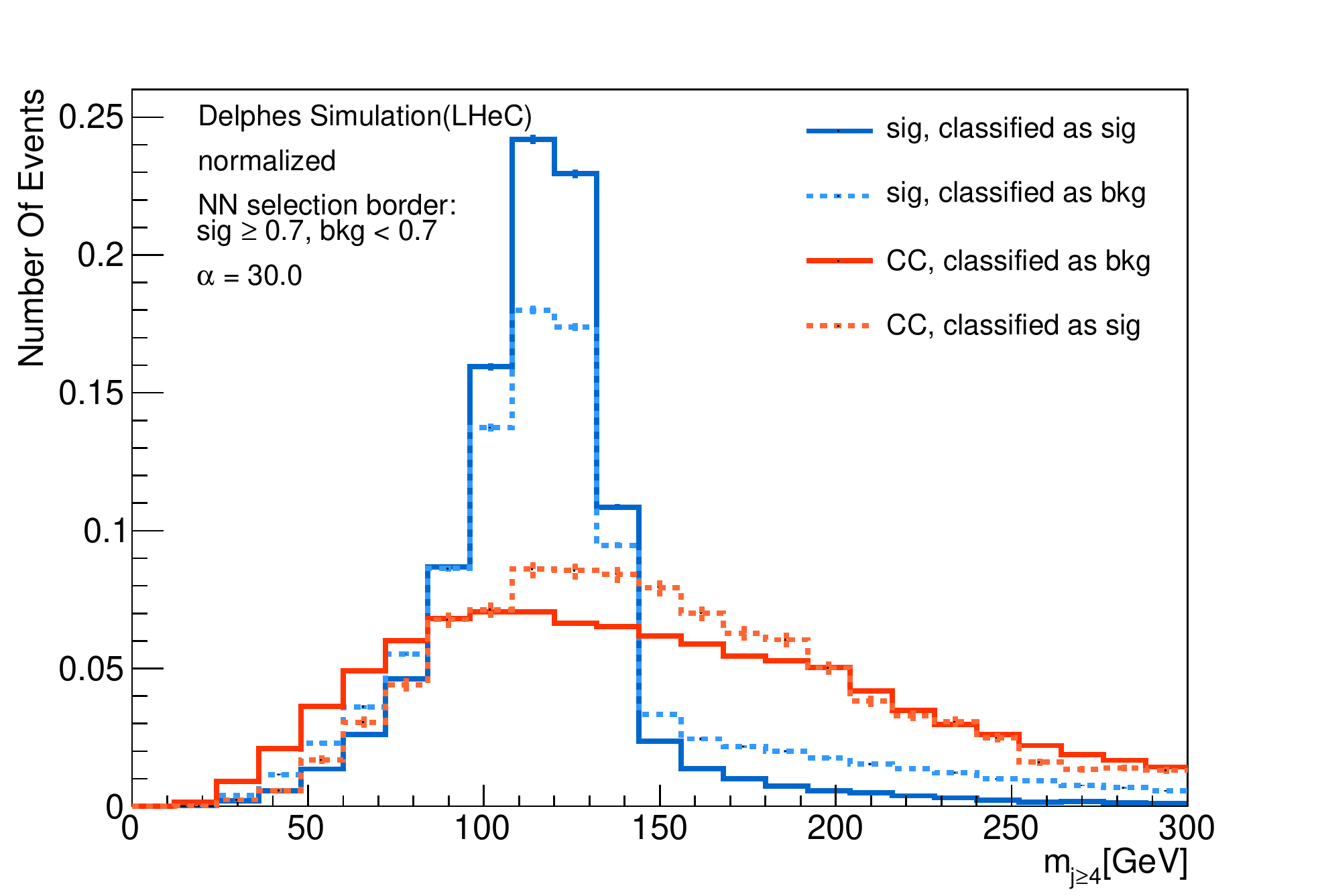}
    \caption{Comparison of the distribution of dominant background processes in the signal and the background dominated region, defined by the NN based classifier for the $H\rightarrow 4j$ with $\alpha = 5$(left) and $\alpha = 30$ (right).}
\label{fig:backgroundshape alpha comparison}
\end{figure}

\begin{figure}[tbp]
\centering
    \includegraphics[width=0.49\textwidth]{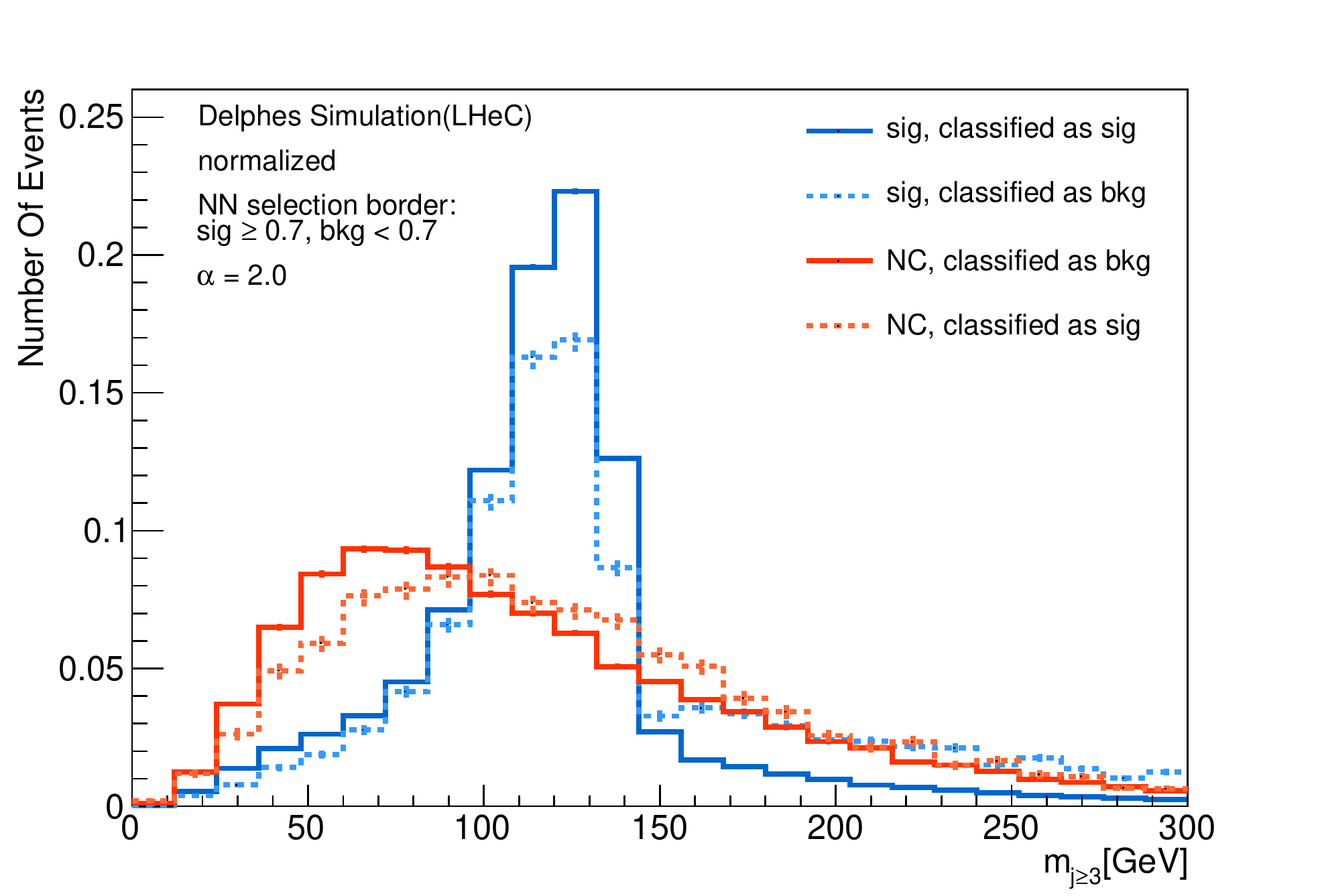}
    \includegraphics[width=0.49\textwidth]{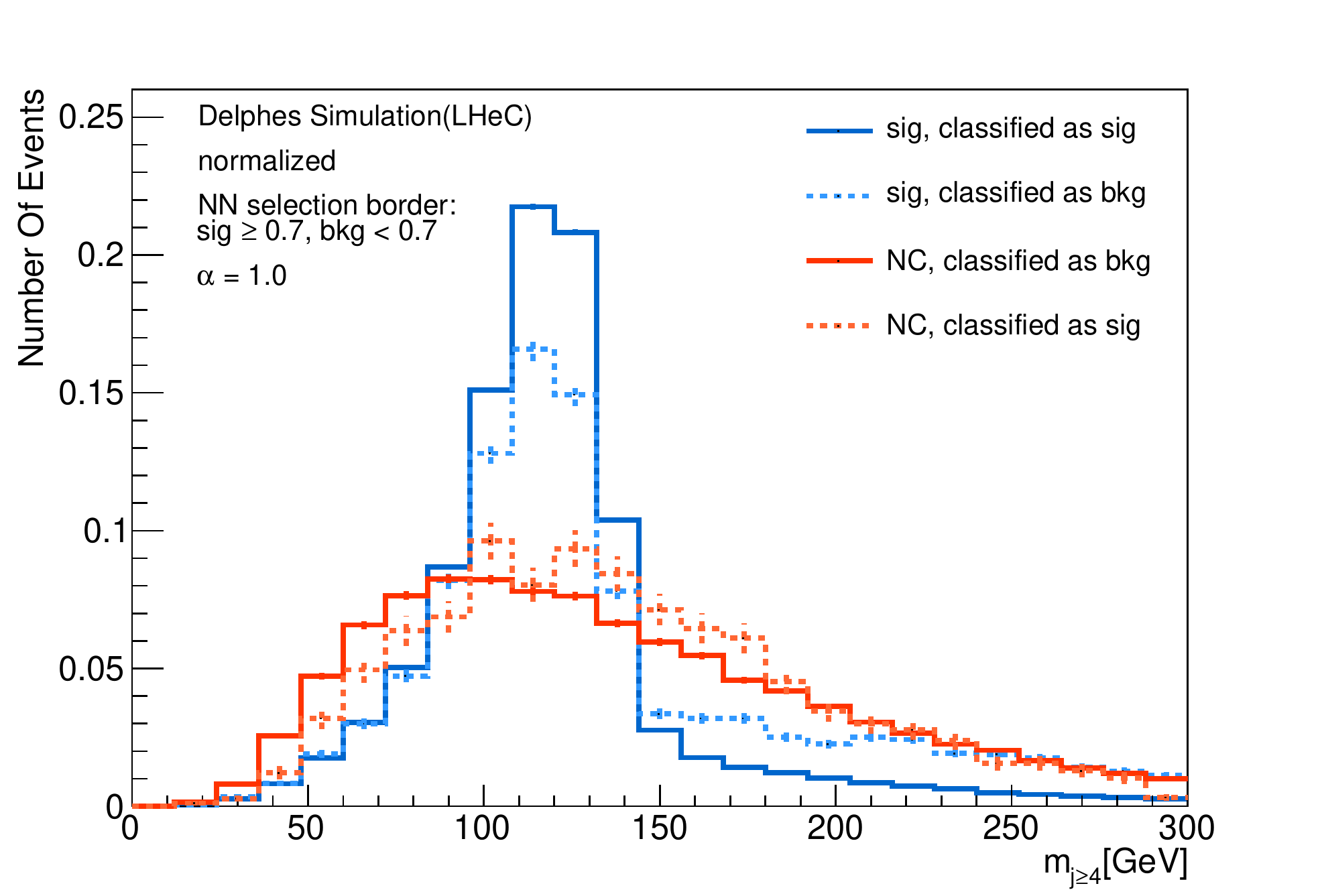}
    \includegraphics[width=0.49\textwidth]{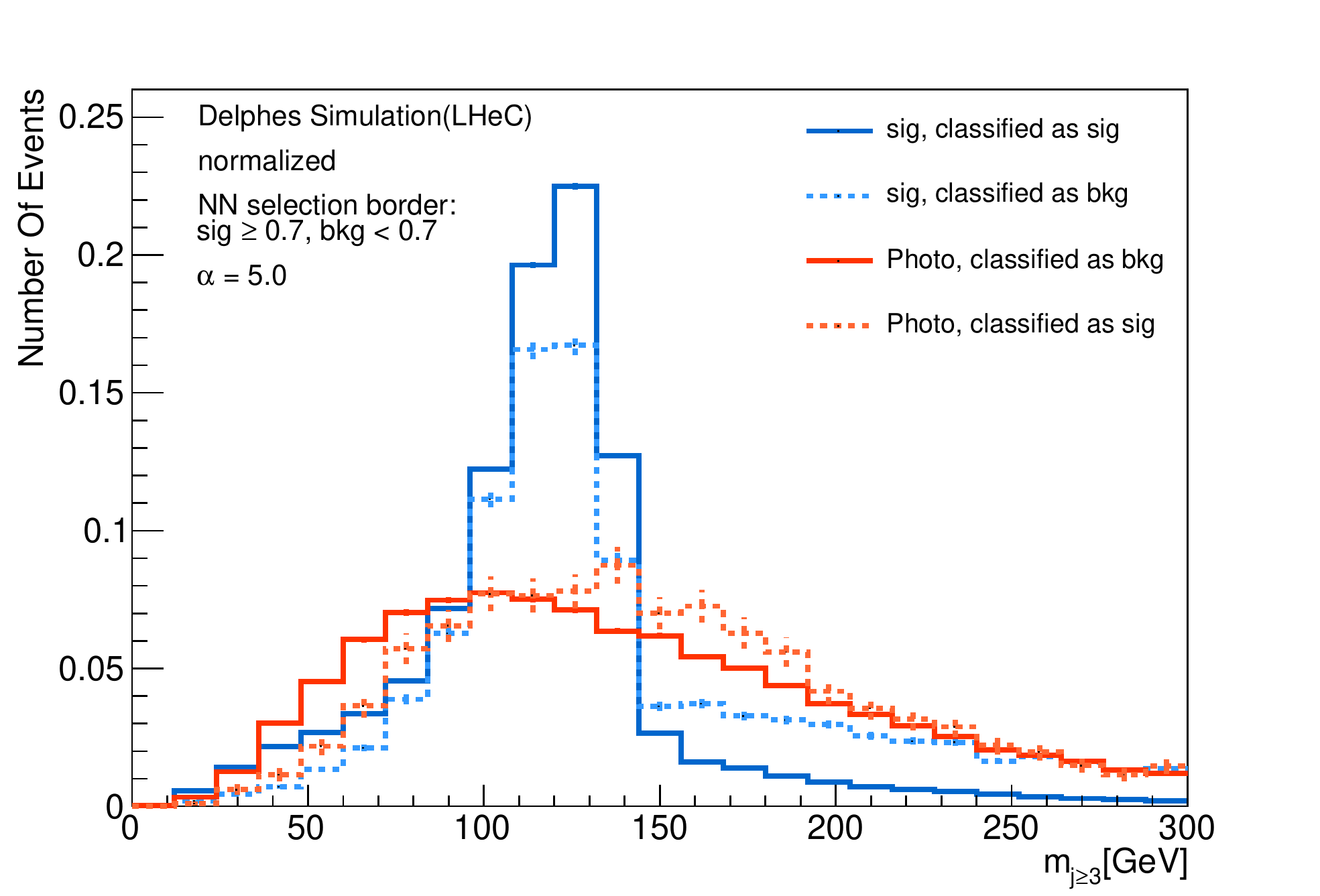}
    \includegraphics[width=0.49\textwidth]{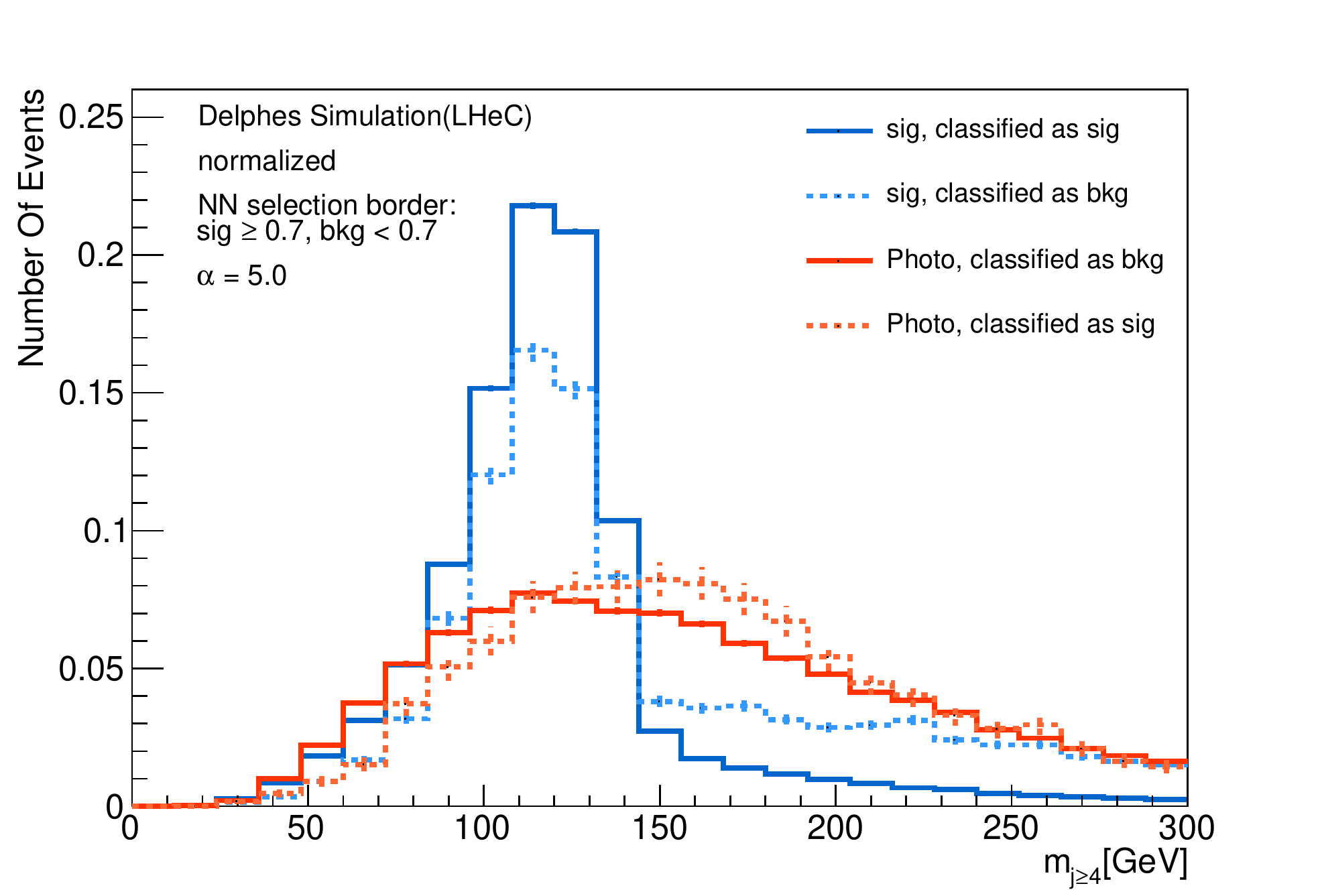}
    \includegraphics[width=0.49\textwidth]{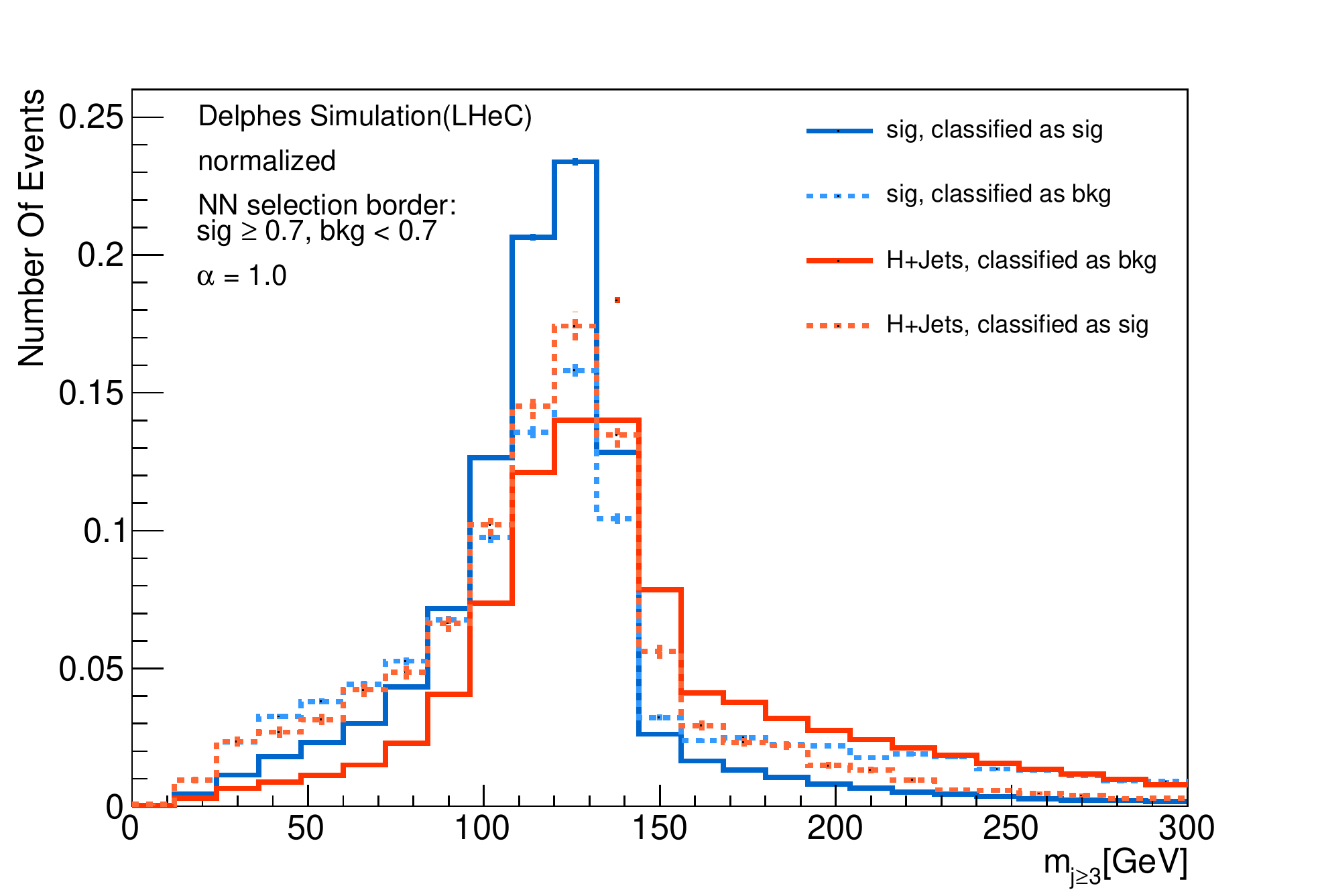}
    \includegraphics[width=0.49\textwidth]{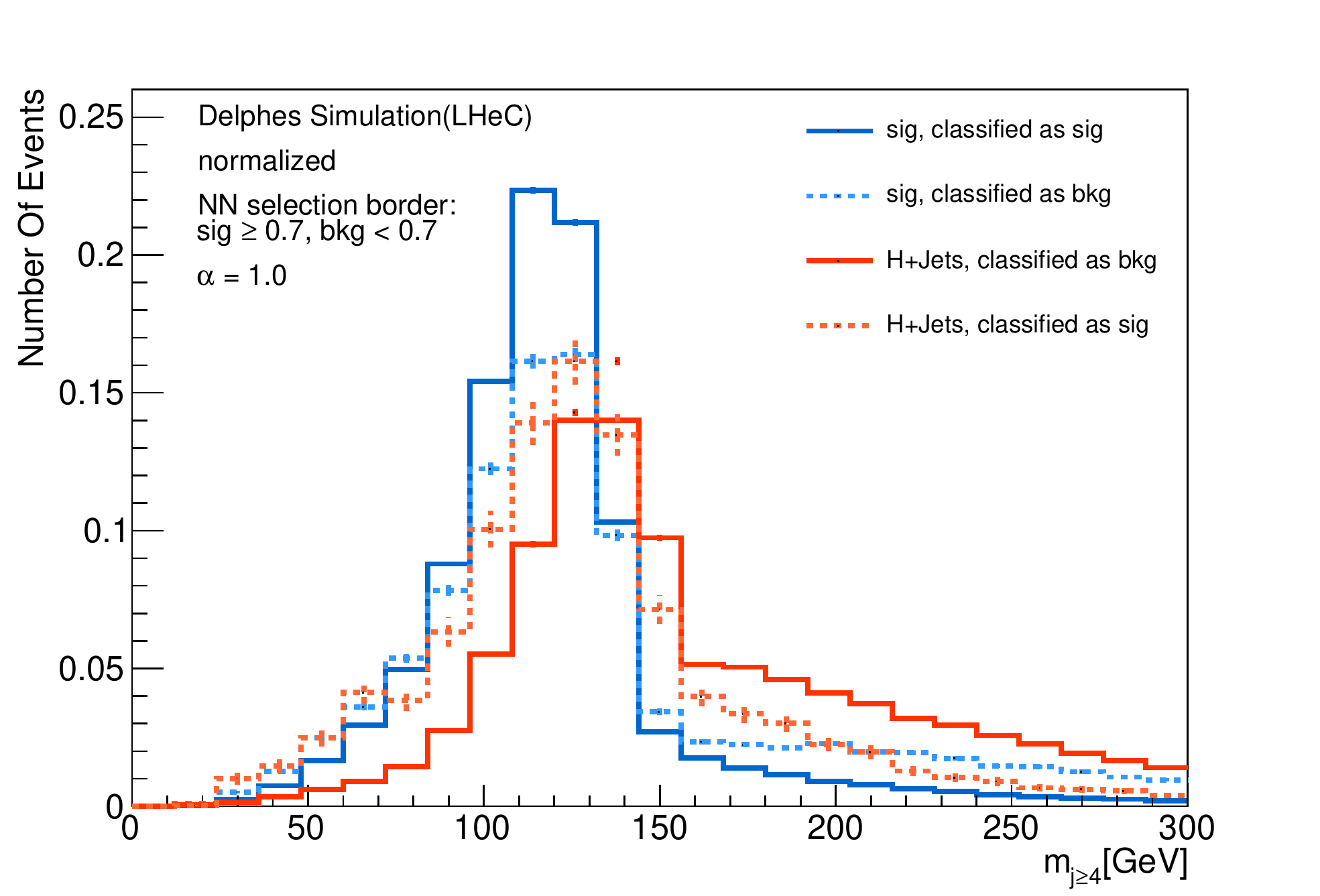}
    \includegraphics[width=0.49\textwidth]{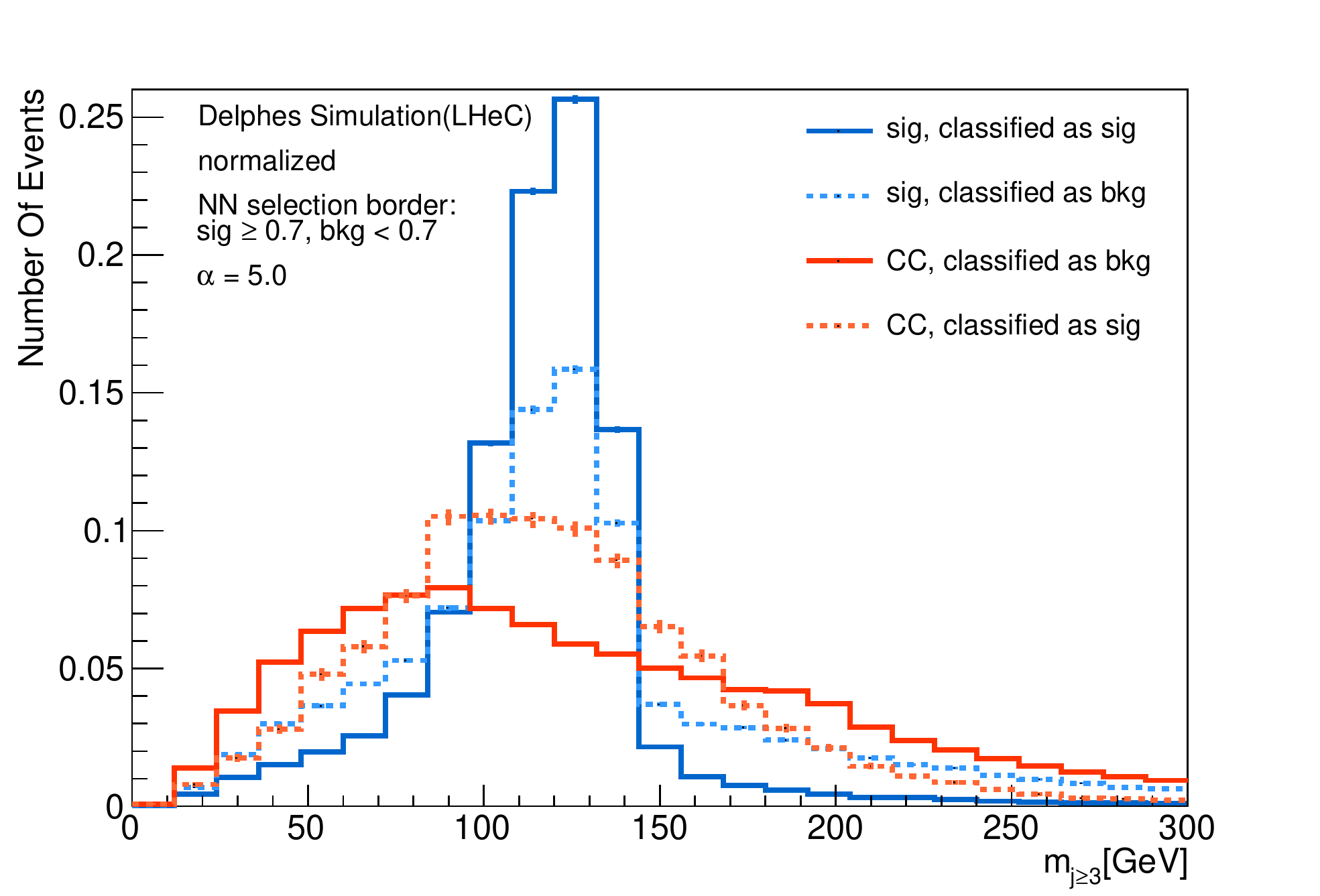}
    \includegraphics[width=0.49\textwidth]{./compare_class_bkg_to_sig_4j_CC_alpha5.pdf}
    \caption{Comparison of the distribution of dominant background processes in the signal and the background dominated region, defined by the NN based classifier for the $H\rightarrow 3j$ case (left) and the $H\rightarrow 4j$ case (right).}
\label{fig:backgroundshape}
\end{figure}

Figure \ref{fig:backgroundshape} presents the shapes of the signals and backgrounds for the four background processes and the $H\rightarrow 3j$ and $H\rightarrow 4j$ signal processes in the signal and background regions. The signal and background regions are defined by applying a threshold value of 0.7 on the classifier with the best-fitting $\alpha$.\\
As depicted in Figure \ref{fig:backgroundshape alpha comparison}, the similarity between the shapes of the signal and background distributions in the signal and background regions increases as the value of $\alpha$ increases. However, a lower value of $\alpha$ results in improved discrimination between the signal and background, although the resulting distributions of $m_{j \geq 3}$ and $m_{j \geq 4}$ after signal selection become increasingly similar to the signal, making the estimation of the background in the side-band regions of the signal more challenging. To balance these factors, a value of $\alpha=5$ is selected for the Photo and CC background processes, $\alpha=1$ for the H+Jets processes, and for NC $\alpha=2$ in the $H\rightarrow 3j$ case, and $\alpha=1$ in the $H\rightarrow 4j$ case.

The threshold values of the four classifiers for the NC, CC, Photo, and H+Jets background processes have been optimized to maximize the ratio of signal events over the square root of the number of background events. An event is considered a signal candidate if all four classifiers produce output values above the respective threshold. Table \ref{tab:cutflow1} and \ref{tab:cutflow2} provide a summary of the cutflows for the $H\rightarrow 3j$ and $H\rightarrow 4j$ signal and background processes, respectively, normalized to an integrated luminosity of $\int L = 1 $ab$^{-1}$, and include only statistical uncertainties. The preselection includes all events that pass the cuts described in Section \ref{sec:Signal}. As expected, these cuts effectively reduce the number of NC and Photo background events. The NN Classification events are those that remain after the Neural Network classification. In the case of the NC and Photo backgrounds, only a few events (around 10) remain, and interpolation on the Neural Network output was used to determine the error in the number of events. The shapes of these background processes were defined using the DisCo property, as described in Section \ref{sec:Signal}.

\begin{table}[ht]
\begin{center}
\begin{tabular}{l | ccccc}
\hline
Cut			& 	Sig H(3j)	& NC    &   Photo   &   H+Jets  &   CC  \\
All events [$\times10^{6}$]     &   $0.16\pm0.001$&   $3079\pm5$ &   $2347\pm4$  &   $0.172\pm0.001$ & $40.6\pm0.1$\\
Preselection [$\times10^{3}$]   &   $136.7\pm0.3$&   $437300\pm1900$  &   $368900\pm1500$  &   $92.6\pm0.2$ &    $26900\pm80$\\
NN Classification		&   $32201\pm161$   &   $271500\pm13068$ &   $30841\pm3734$  &   $1430\pm30$ &  $141563\pm5496$\\
\hline
\end{tabular}
\end{center}
\caption{Cut-Flow for "higgs particle decaying into 3 jet"-signal and background samples corresponding to an integrated luminosity of $\int L = 1 ab^{-1}$.}
\label{tab:cutflow1}
\end{table}

\begin{table}[ht]
\begin{center}
\begin{tabular}{l | ccccc}
\hline
Cut			& 	Sig H(4j)	& NC    &   Photo   &   H+Jets  &   CC  \\
All events [$\times10^{6}$]     &   $0.16\pm0.001$&   $3079\pm5$ &   $2347\pm4$  &   $0.172\pm0.001$ & $40.6\pm0.1$\\
Preselection [$\times10^{3}$]   &   $135.7\pm0.3$&   $271000\pm1500$  &   $286400\pm1300$  &   $45.8\pm0.1$ &    $16060\pm60$\\
NN Classification		&   $36684\pm171$   &   $52754\pm4402$ &   $16652\pm1482$  &   $281\pm13$ &  $40467\pm2738$\\
\hline
\end{tabular}
\end{center}
\caption{Cut-Flow for "higgs particle decaying into 4 jet"-signal and background samples corresponding to an integrated luminosity of $\int L = 1 ab^{-1}$.}
\label{tab:cutflow2}
\end{table}

\section{Expected Sensitivity\label{sec:Background}}

\begin{figure*}[thb]
\centering
    \includegraphics[width=0.49\textwidth]{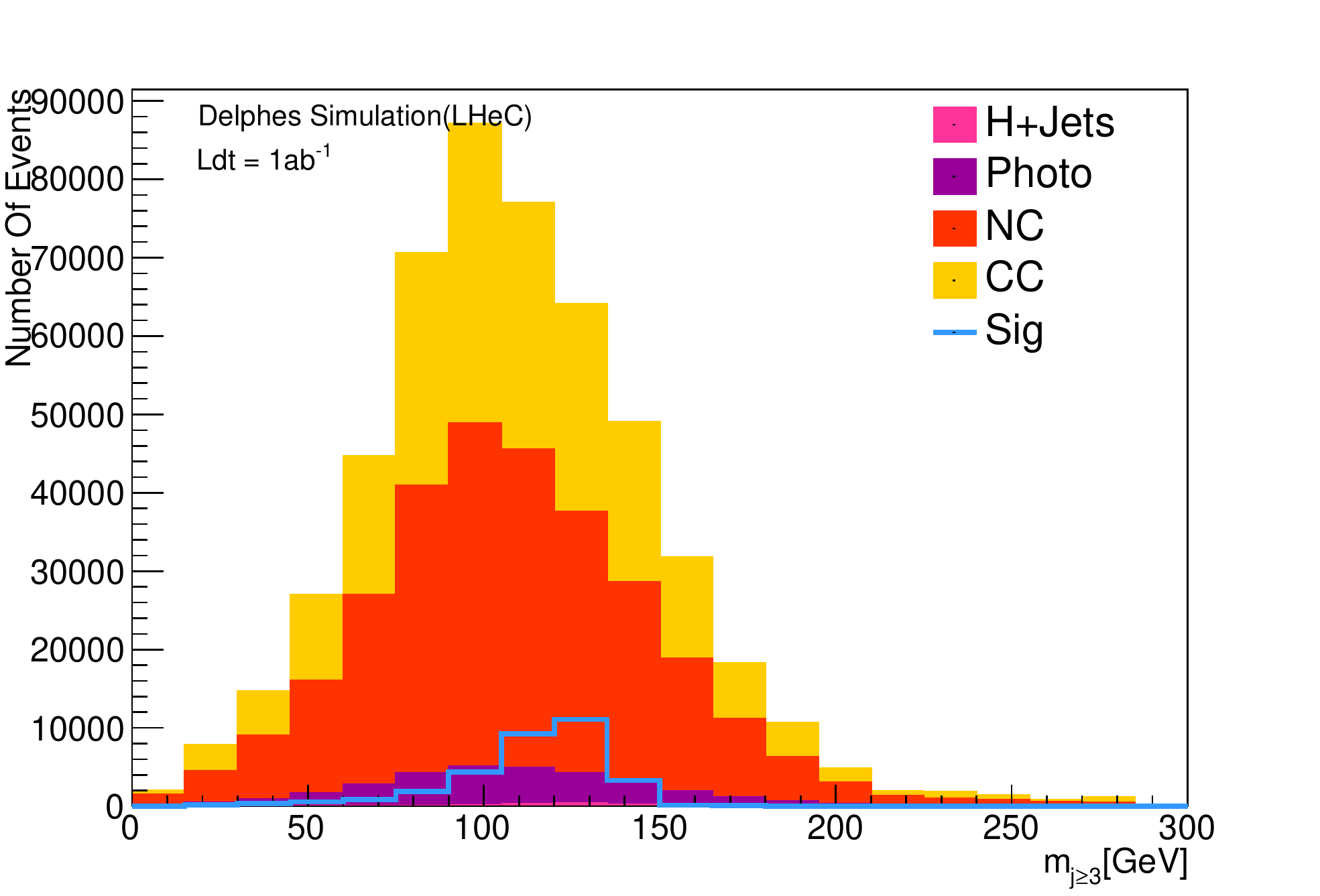}
    \includegraphics[width=0.49\textwidth]{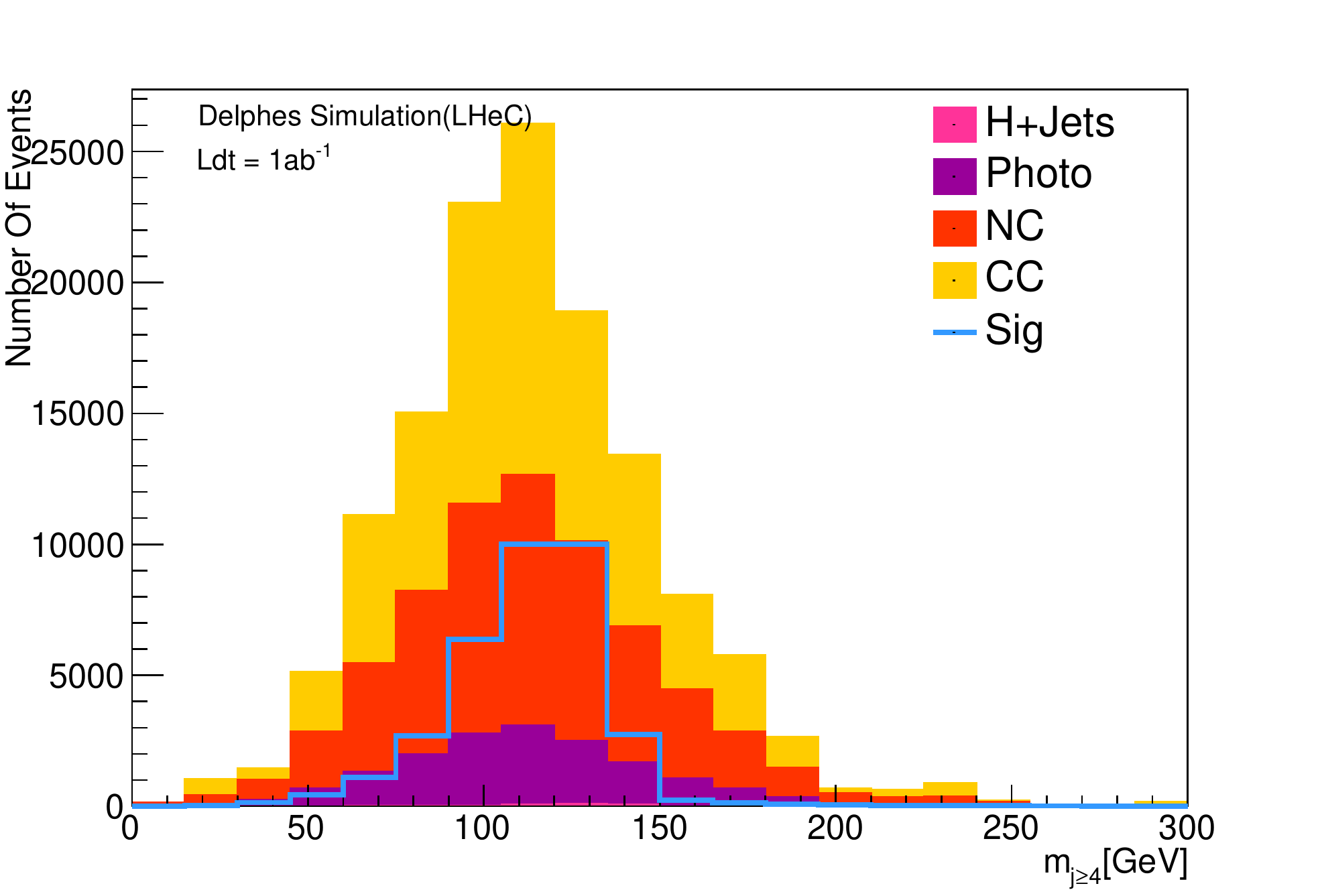}
    \caption{Expected invariant mass distribution of selected jets after the full signal selection for an integrated luminosity of $\int L = 1 ab^{-1}$ for the $H\rightarrow 3j$ (left) and the $H\rightarrow 4j$ case (right).}
\label{fig:invariantmass}
\end{figure*}

The expected distributions of $m_{j \geq 3}$ and $m_{j \geq 4}$ after full event selection for both signal and background processes are displayed in Figure \ref{fig:invariantmass} for an integrated luminosity of $\int L = 1\text{ ab}^{-1}$. A distinct difference in the shapes of the signal and background distributions can be observed. The regions below and above the Higgs boson peak can be utilized to normalize the various background processes and facilitate constraints on any potential signals.

\begin{figure*}[thb]
\centering
    \includegraphics[width=0.49\textwidth]{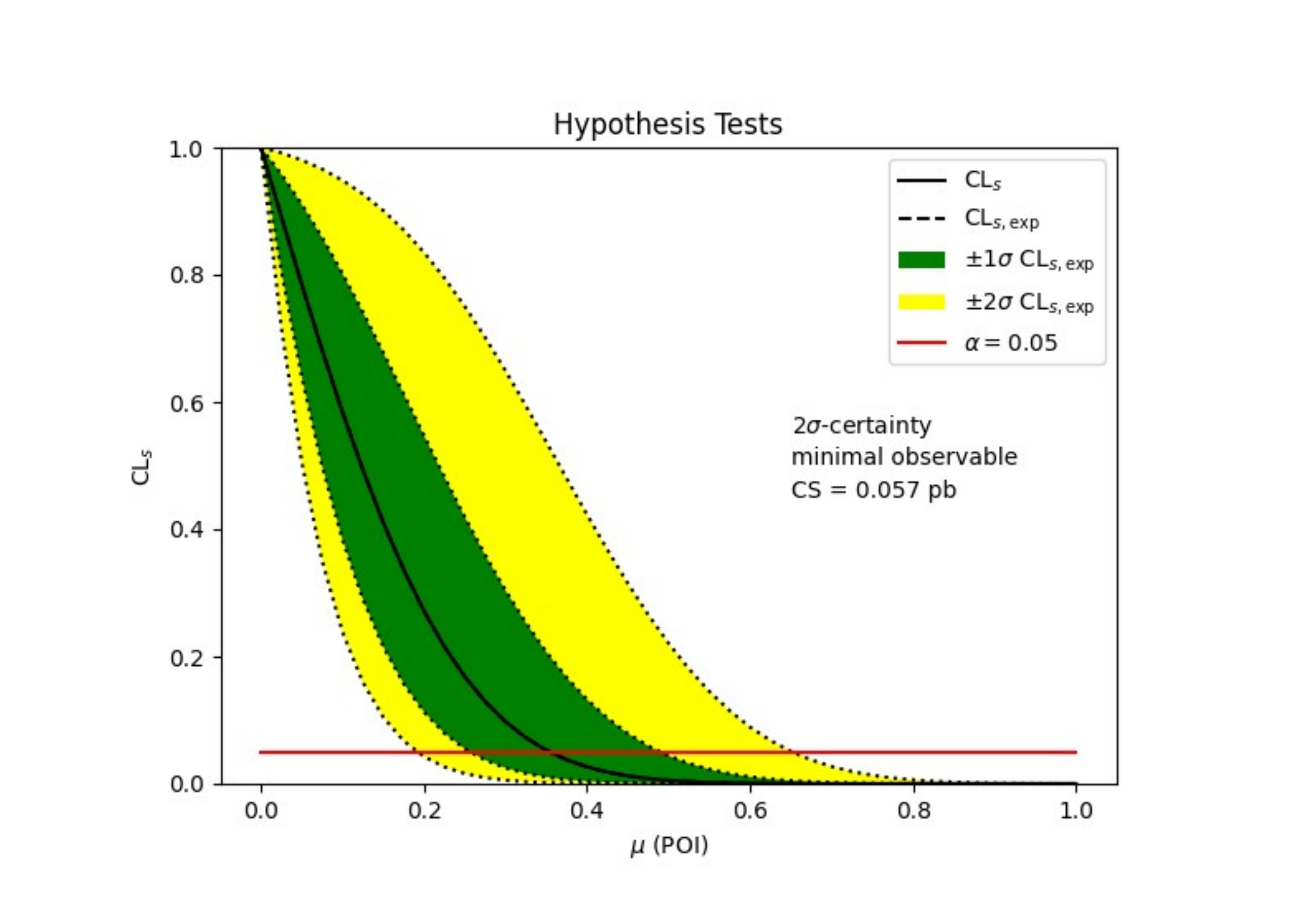}
    \includegraphics[width=0.49\textwidth]{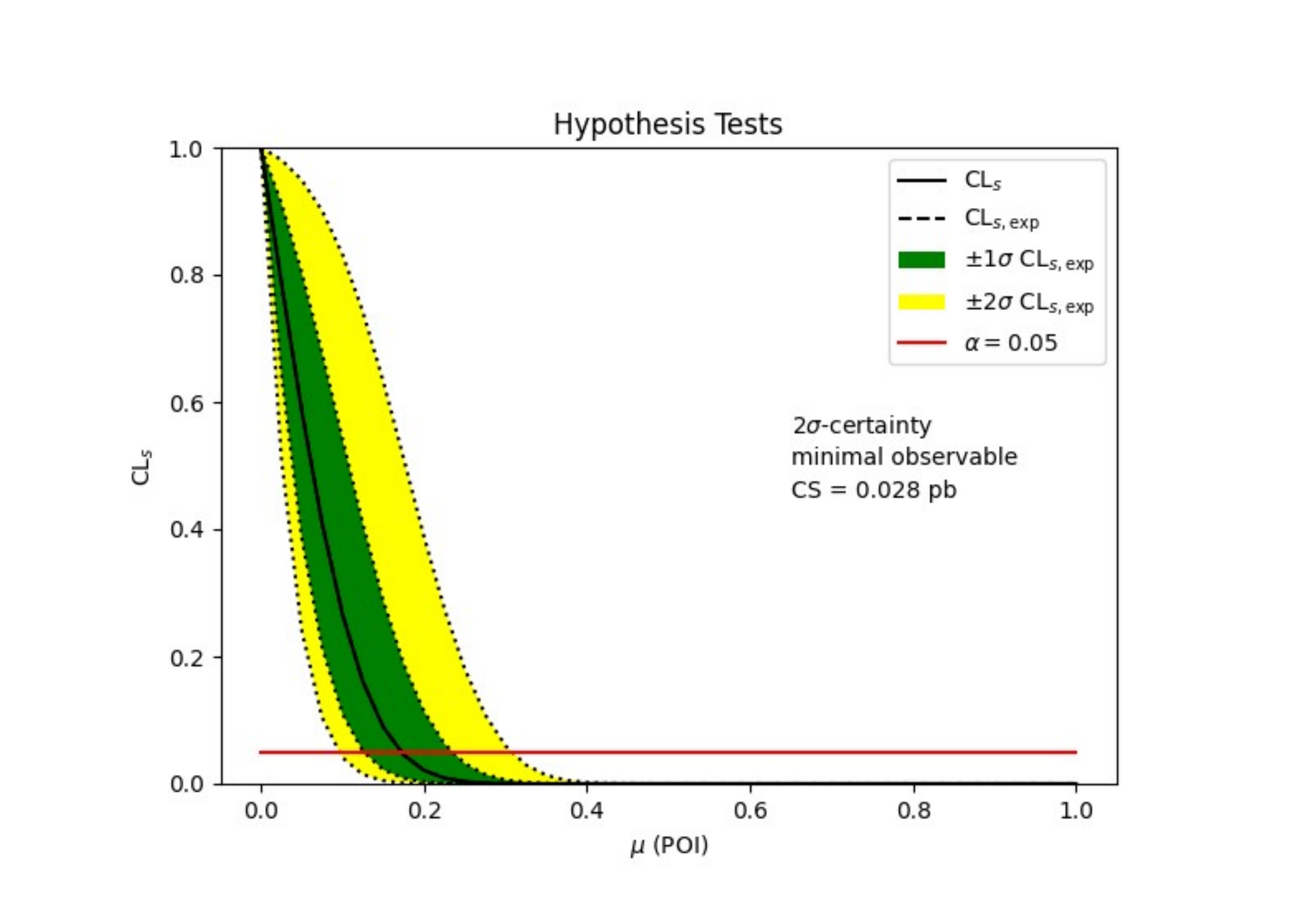}
    \caption{Expected exclusion limit for $H\rightarrow 3j$ (left) and the $H\rightarrow 4j$ for an integrated luminosity of $\int L = 1 ab^{-1}$, where the x-axis shows the signal-strength parameter $\mu$ that scales the signal cross-section. A value of $\mu=1$ corresponds to the Standard Model Higgs boson production cross section at the LHeC.}
\label{fig:limit}
\end{figure*}

To estimate the expected sensitivities of the analysis, the number of background events selected as signal in each mass bin is used to generate toy data sets via random variation based on a Poisson distribution. The statistical analysis is performed using the PyHF framework \cite{Heinrich:2021gyp}, which employs a maximum-likelihood fit to the $m_{j \geq 3}$ and $m_{j \geq 4}$ distributions. The hypothesis of background-only compatibility with the toy data is tested using a profile-likelihood-ratio test statistic in the absence of anomalous Higgs boson decays. The signal cross-section is assumed to equal the expected SM cross-section of Higgs boson production at the LHeC via CC processes, i.e. $\sigma(pe^- \to \nu_e H q)= 0.160$ pb for the beam energy $(E_p,~E_e)$=(7000, 60) GeV with 80\% left polarized initial electron beam. The upper limits at 95\% CL on the signal strength parameter $\mu$, which scales the signal cross section, are derived separately for the $H\rightarrow 3j$ and $H\rightarrow 4j$ processes and are shown in Figure \ref{fig:limit}. Without taking into account systematic uncertainties, the smallest observable signal cross section is expected to have an upper limit of $CS_{H \rightarrow 3j} = 0.056$ pb and $CS_{H \rightarrow 4j} = 0.025$ pb, respectively, corresponding to an upper limit on the branching ratio $BR(\sigma_{ep\rightarrow H+X\rightarrow 3/4j+X}/\sigma_{ep\rightarrow H+X})$ of 0.35 and 0.15. 

In a second step, we also estimate the impact of potential systematic uncertainties, in particular, a jet energy scale and jet energy resolution uncertainty in the order of 1\% and 10\% respectively. This induces correlated uncertainties in the expected signal and background shapes in the $m_{j \geq 3}$ and $m_{j \geq 4}$ distributions. Repeating the fit with these additional uncertainties yields upper limits of 0.35 and 0.17 for both cases, with reduced sensitivity of about 0.5\% and 11\% respectively. 

\section{Conclusion\label{sec:Conclusion}}

While many possible decay channels of the Higgs boson can be studied with high precision at the LHC, hadronic Higgs boson decays are extremely challenging due to the large multi-jet cross sections at the LHC.  In this work, we estimated the prospects of searches for anomalous hadronic Higgs boson decays, in particular, $H\rightarrow 3j$ and $H\rightarrow 4j$ processes at the LHeC with an integrated luminosity of ${\it \int L dt} = 1 ab^{-1}$. We expect upper limits on the BR of 0.35 and 0.17 at 95\% confidence limit for those processes, respectively. Our studies should be seen only as a starting point for more detailed studies on Higgs hadronic final states at the LHC, maybe with the final goal to search for two jet and multi-hadron final state signatures.  

\section*{Acknowledgement}
This proposal would not have been possible without the support of the PRISMA$^+$ Cluster of Excellence as well as the DFG Grant SCHO 1527/6-1. 
	
\bibliographystyle{elsarticle-num}
\bibliography{Bibliography}

\end{document}